\documentclass[journal]{IEEEtran}

\usepackage{cite}
\usepackage{amsmath,amssymb,amsfonts}
\usepackage{graphicx}
\usepackage{textcomp}
\usepackage{xcolor}
\usepackage{enumitem}
\usepackage{algorithm}
\usepackage{algpseudocode}
\usepackage{verbatim}
\usepackage{amsthm}
\newtheorem{theorem}{Theorem}
\newtheorem{lemma}{Lemma}

\usepackage{url}
\usepackage{bm}
\usepackage[nocomma]{optidef}
\newcommand{\thing}{appendix_on}
\newcommand{\lemmazero}{yes}
\newcommand{\lemmaone}{yes}
\newcommand{\lemmatwo}{yes}
\newcommand{\lemmathree}{yes}
\newcommand{\lemmafour}{yes}
\newcommand{\lemmafive}{yes}
\newcommand{\lemmasix}{yes}
\usepackage{ifthen}
\usepackage{hyperref}
\usepackage{cleveref} 
\crefname{lemma}{Lemma}{Lemmas}
\usepackage{ifpdf}
\usepackage{subcaption}
\ifCLASSINFOpdf
\else
\fi

\ifthenelse{\equal{\thing}{appendix_on}}
  {
   \newcommand{\result}{\onecolumn
\appendix
\label{sectionappendix}

   We prove Theorem \ref{theorem_feasibility_optimal} towards the end of this section.  We initially prove some results which help prove Theorem \ref{theorem_feasibility_optimal}.
    Our approach in this proof is an extension of the approach in \cite{prk} (which does not consider bandwidth sharing) and uses Lyapunov optimization techniques such as those in \cite{neely}.
 
 \subsection{Notation and simplifications}
 Recall that $Q^{i}_{r,n}(\tau^{i}_{r,n}(k))$ is specific for each client  in every region of every operator. For simplifying the notation in this section, we denote $Q^{i}_{r,n}(\tau^{i}_{r,n}(k))$ as $Q(\tau^{i}_{r,n}(k))$ in this section.
 Vector notation used in this section is listed in Table \ref{proof notations}.
\begin{table}[htp]
\caption{Vector notation used in this section}
    \centering
    \begin{tabular}{p{0.05\textwidth} p{0.4\textwidth}}
    \hline
\textbf{Notation}  & \textbf{Description} \\
\hline
\hline
$\bm{q}$ & Array with entries $q_{r,n}^{i}$  for each $n\in {\mathcal N}^{i}_{r}, \ \forall{r} \in {\mathcal R}, \  \forall {i} \in {\mathcal O} $\\
 \\[-1em]
$\bm{\zeta}$ &  Array with entries $\zeta^{(i,j)}$  for each ${i,j} \in \mathcal O $\\
 \\[-1em]
 $\bm{\tau}(k)$ & Array with entries $\tau_{r,n}^{i}(k)$ for each $n\in {\mathcal N}^{i}_{r}, \ \forall{r} \in {\mathcal R}, \  \forall {i} \in {\mathcal O}$\\
  \\[-1em]
 $\bm{\delta}(k)$ & Array with entries $\delta_{r,n}^{i}(k)$ for each $n\in {\mathcal N}^{i}_{r}, \ \forall{r} \in {\mathcal R}, \  \forall {i} \in {\mathcal O}$\\
 \\ [-1em]
 $\bm{\sigma}(k)$ & Array with entries $\sigma^{i\rightarrow j}(k)$  for each ${i,j} \in \mathcal O$\\
         \hline
         \hline
    \end{tabular}
    \label{proof notations}
\end{table}

Let $\Theta^*(k) = \left(\bm{\delta}^*(k),\bm{\sigma}^*(k)\right)$ track the virtual queues under our policy evolving according to \eqref{debts1} and \eqref{debts2}.
Define quadratic Lyapunov function $L(\Theta^*(k)$) as
\begin{align}
  &  L(\Theta^*(k))=\frac{1}{2} \left [ \sum_{r \in \mathcal {R}} \sum_{i\in \mathcal O}\sum_{{n} \in {\mathcal N}^{i}_{r}}\delta_{r,n}^{*i}(k)^2 +\sum_{i\in \mathcal O}\sum_{j\in {\mathcal O}\setminus\{i\}}\sigma^{*i\rightarrow j}(k)^2\right], \qquad \forall { k \in \mathcal K}.\label{eqn:lf}
\end{align}

We consider stationary randomized policies in the next subsection, which take the actions in a randomized manner based solely on the packet arrivals in the period (i.e., the "state" in the period).
Let ${\tau}_{r,n}^{**i}(k)$ denote the number of timeslots allotted by an asymptotically optimal stationary randomized policy.
Let $Q\left({\tau}_{r,n}^{**i}(k)\right)$ denote the corresponding perceived quality.
Let $U^{**}= \sum_{r \in \mathcal {R}} \sum_{i\in \mathcal O}\sum_{{n} \in {\mathcal N}^{i}_{r}} {E\left[Q\left(\tau_{r,n}^{**i}(k)\right) \right ]}$, i.e., the optimal value of \eqref{objfunctionexpression} obtained by an asymptotically optimal stationary randomized policy.
Further, let
\begin{eqnarray}
   U(\bm{\tau}^*(k))) &=& \sum_{r \in \mathcal {R}} \sum_{i\in \mathcal O}\sum_{{n} \in {\mathcal N}^{i}_{r}}Q(\tau_{r,n}^{*i}(k)). \label{UQ*}
 \end{eqnarray}

\subsection{Intermediate results}

 We start by stating and proving that the control actions taken under a stationary randomized policy (\cref{b rate convergent}) are rate convergent if the arrival process is rate convergent.

\begin{lemma}\label{b rate convergent}
    The processes $\ddot{b}_{r,n}^{i}(k)$ and $\ddot{S}^{j\rightarrow i}_{r}(k)$ that denote the acceptable video quality indicator and  the number of timeslots shared respectively, under any stationary randomized policy are rate convergent if the arrival process $A_{r,n}^{i}(k)$ is rate convergent.
\end{lemma}

\lemmazeroproof

The next lemma is essentially a condition on the drift of the Lyapunov function defined in \eqref{eqn:lf}.
\begin{lemma}\label{drift- quality lemma}
 [\textbf{$\bm{M}$-step drift-quality condition}]  There exist  positive constants $C>0$, $\alpha >0$ and $\beta >0$ such that for each period $k$,
 \begin{align}
 E\left[ \frac{L(\Theta^*(k+M))}{M}-\frac{L(\Theta^*(k))}{M} -\frac{V}{M}\sum_{x=k}^{k+M-1} U(\bm{\tau}^*(x))) \bigg|\Theta(k)\right ] \leq  C  & - V U^{**}-\alpha \sum_{r \in \mathcal {R}} \sum_{i\in \mathcal O}\sum_{{n} \in {\mathcal N}^{i}_{r}(k)}\delta_{r,n}^{*i}(k) \nonumber \\
& - \beta \sum_{i\in \mathcal O}\sum_{j\in {\mathcal O}\setminus\{i\}}\sigma^{*i\rightarrow j}(k)\label{lemma1ineq}
\end{align}
\end{lemma}

 \lemmaoneproof

The next result is an important intermediate result proved using \cref{drift- quality lemma}, upper bounding on time average of quality debts and sharing debts, and lower bounding the value of objective function \eqref{objfunctionexpression} under our policy.

\begin{lemma}\label{average queue lemma}
There exist positive constants $C>0$, $\alpha >0$ and $\beta >0$, such that for any positive integer $K$
\begin{align}
& \alpha \frac{1}{K}\sum_{k=0}^{K-1} E \left [\sum_{r \in \mathcal {R}} \sum_{i\in \mathcal O}\sum_{{n} \in {\mathcal N}^{i}_{r}(k)}\delta_{r,n}^{*i}(kM)\right ] + \beta \frac{1}{K}\sum_{k=0}^{K-1}  E \left[\sum_{i\in \mathcal O}\sum_{j\in {\mathcal O}\setminus\{i\}}\sigma^{*i\rightarrow j}(kM)\right] \nonumber \\
&\leq C + \frac{1}{K} \sum_{k=0}^{K-1} \frac{V}{M}  E \left[\sum_{x=kM}^{kM+M-1} U(\bm{\tau}^*(x))\right] - V U^{**} -\frac{ E [ \hat{L}(\Theta^*(K)) ]}{K}+ 
\frac{E [\hat{L}(\Theta^*(0))]}{K} 
 \end{align}
 \end{lemma}

 \lemmatwoproof

The following result states that the time average expected quality under our policy can be made arbitrarily close to the optimal value $U^{**}$ by choosing a suitably large $V$.

\begin{lemma}\label{average quality lemma}
 There exists 
 positive constant $C>0$, such that
\begin{align}
& \liminf_{K \to \infty} \frac{1}{K} \sum_{r \in \mathcal {R}} \sum_{i\in \mathcal O}\sum_{{n} \in {\mathcal N}^{i}_{r}} \sum_{k=0}^{K-1} \frac{1}{M} \sum_{x=kM}^{kM+M-1} E \left[ Q(\tau_{r,n}^{*i}(x))\right]
\geq   U^{**} -\frac{C}{V}
\end{align}

 \end{lemma}

 \lemmafourproof
 
 Following is a result from \cite{prk} reproduced below for easy reference.
\begin{lemma} \textbf{(from \cite{prk})} \label{prk_lemma}
 Let $f(n)$ be a nonnegative function such that $|f(n+1)-f(n)| \leq M$, for some $M>0$, for all n. \\
 If $\limsup_{n \to \infty}\frac{1}{n}\sum_{i=0}^n f(i) \leq B, \text{~for some constant B, then~ }\lim_{n \to \infty}\frac{1}{n}f(n)=0$ 
\end{lemma}

Now, we have the following result.
\begin{lemma}\label{lemma_new}
For any strictly feasible timely delivery rate  requirement $\bm{q}$ and sharing bound $\bm{\zeta}$,  we have

\begin{equation}
    \text{Prob}\bigg\{\frac{\delta_{r,n}^{*i}(K)}{K}< \xi_1 \bigg\} \rightarrow 1, \text{~as~} K \rightarrow \infty, \ \forall {{n} \in {\mathcal N}^{i}_{r}(k)}, i\in \mathcal O, r \in \mathcal R, \xi_1 > 0;\label{lemma4_xi1}
\end{equation}
\begin{equation}
    \text{Prob}\bigg\{\frac{\sigma^{*i\rightarrow j}(K)}{K}< \xi_2 \bigg\} \rightarrow 1, \text{~as~} K \rightarrow \infty, \ \forall  i\in \mathcal O, {j\in {\mathcal O}\setminus\{i\}}, r \in \mathcal R, \xi_2 > 0. \label{lemma4_xi2}
\end{equation}
\end{lemma}

\lemmafiveproof

\begin{lemma}\label{1 3 constraint satisfy lemma}
For any strictly feasible timely delivery rate requirement $\bm{q}$ and sharing bound $\bm{\zeta}$, 
\eqref{throughput constraint} and \eqref{sharing constraint} are satisfied.
\end{lemma}

\lemmasixproof

\subsection{Proof of  Theorem \ref{theorem_feasibility_optimal}}

Theorem \ref{theorem_feasibility_optimal} follows from \cref{average quality lemma} and \cref{1 3 constraint satisfy lemma}.

}%
  }
  {
   \newcommand{\result}{}%
  }

 \ifthenelse{\equal{\lemmazero}{yes}}
  {
  }
  {
  }
  
\ifthenelse{\equal{\lemmaone}{yes}}
  {
  }
  {
  }

  \ifthenelse{\equal{\lemmatwo}{yes}}
  {
  }
  {
  }
  
    \ifthenelse{\equal{\lemmathree}{yes}}
  {
  }
  {
  }

    \ifthenelse{\equal{\lemmafour}{yes}}
  {
  }
  {
  }
  
  \ifthenelse{\equal{\lemmafive}{yes}}
  {
  }
  {
  }  

   \ifthenelse{\equal{\lemmasix}{yes}}
  {
  }
  {
  }  

\begin{document}

\title{Maximizing Real-Time Video QoE via 
\\Bandwidth Sharing under Markovian setting}

\author{Sushi~Anna~George, Vinay~Joseph
\thanks{S. George and V. Joseph are with the National Institute of Technology, Calicut, India.}
}


\maketitle

\begin{abstract}
We consider the problem of optimizing Quality of Experience (QoE) 
of clients streaming real-time video, served by networks managed by different operators that can share bandwidth with each other. 
The abundance of real-time video traffic is evident in the popularity of applications like video conferencing and video streaming of live events, which have increased significantly since the recent pandemic. 
We model the problem as a joint optimization of  resource allocation for the clients and bandwidth sharing across the operators, with special attention to how the resource allocation impacts clients' perceived video quality. 
We propose an online policy as a solution, which involves dynamically sharing a portion of one operator's bandwidth with another operator. 
We provide strong theoretical optimality guarantees for the policy. 
We also use extensive simulations to demonstrate the policy's substantial performance improvements (of up to ninety percent), and identify insights into key system parameters (e.g., imbalance in arrival rates or channel conditions of the operators) that dictate the improvements. 
\end{abstract}


\IEEEpeerreviewmaketitle

\section{Introduction}

\IEEEPARstart{T}{his} paper focusses on wireless networks serving real-time video. A key characteristic of real-time video (and real-time traffic in general), is that a packet \textit{has to be delivered within a deadline}, failing which the packet becomes useless (or much less useful). 
Real-time traffic is key to several applications like video and audio conferencing, video streaming of live events, augmented and virtual reality applications, and online gaming. 
The relevance of real-time traffic increased substantially after the start of the COVID-19 pandemic. 
Video-conferencing generates a substantial volume of real-time traffic, and its market was predicted to grow by 12\% between 2018-2023 \cite{RM}. The pandemic boosted this market, which is now expected to grow at the rate of 23\% till 2027 \cite{gmi}. 
\textcolor{black}{Also, live video streaming is predicted to grow annually at the rate of 22.4\% from 2021 to 2028 \cite{meti_research}.}

To tackle the projected real-time traffic growth, we develop a solution based on \textit{bandwidth sharing} to better support real-time traffic in wireless networks.
Bandwidth sharing or resource pooling itself is not a novel concept (see more discussion in Section \ref{section_related_work}).
However, it is worthwhile to note here that \textit{bandwidth sharing is particularly attractive for real-time traffic}, as the traffic has to be served within a tight timeline. 
\textcolor{black}{We discuss this using classic queuing theory results in Section \ref{section_related_work}.}

\begin{figure}[htbp]
\centerline{\includegraphics[scale=0.25]{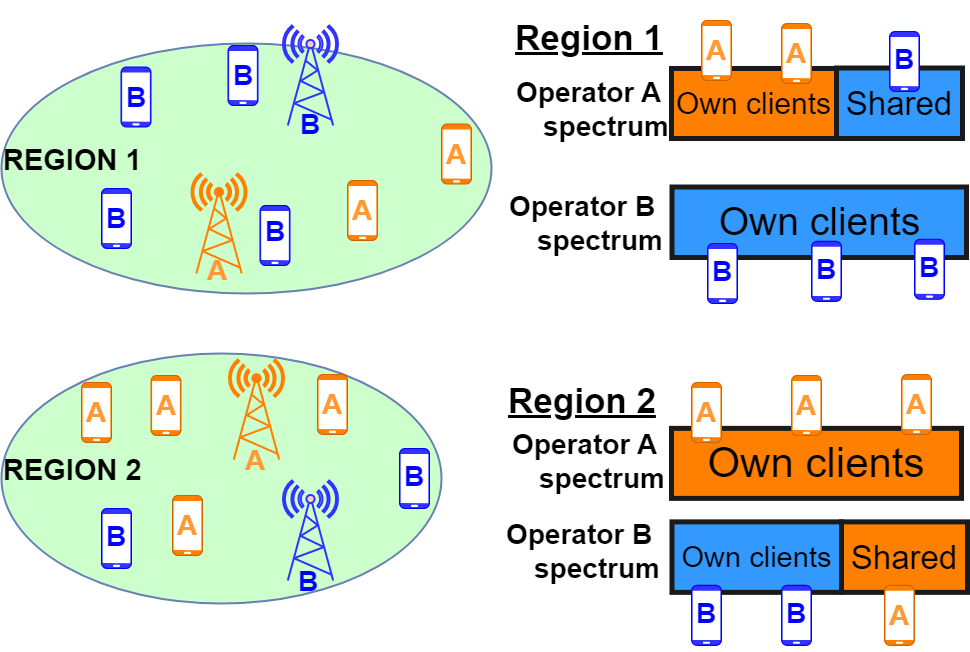}}
\caption{Illustration of bandwidth sharing between operators across regions}
\label{sharingillustration}
\end{figure}

In this paper, we consider dynamic wireless bandwidth sharing between  operators across multiple geographical regions. The basic idea is illustrated in Fig. \ref{sharingillustration}. 
The bandwidth sharing mechanism enables an operator $A$ with lower real-time traffic in region 1 
to share some of its timeslots with another operator $B$ in the same region. 
 \textcolor{black}{Bandwidth sharing can also be useful if there is an imbalance in channel conditions of the 
 operators, like when one operator has lower band spectrum (e.g., 1 GHz)  and the other has higher band spectrum (e.g., 3.5 GHz).}
 
Bandwidth sharing requires strong incentives for the operators who are likely competitors competing for the same customers. Thus, we consider a sharing framework where the net amount of bandwidth shared between any two operators across all regions is forced to be equal. That is, even if operator $A$ shares more bandwidth with operator $B$ in region 1, the sharing framework requires that operator $B$ compensates for this by sharing more in other regions. The operators in this case may be using licensed spectrum, in which case sharing can be realized using techniques like carrier aggregation \cite{3gpp_carrier}. 
The operators could alternatively be using unlicensed spectrum in a coordinated manner in colocated deployments (e.g., using NR-Unlicensed), where different operators coordinate to use different parts of the unlicensed spectrum (to avoid interference) for their regular operation and use sharing on top of it.

Bandwidth sharing or spectrum sharing is being actively discussed in many forums across the world including government agencies, industry and other stakeholders. UK's telecom regulator Ofcom (see \cite{ofcom}) and Communications Technology Laboratory of the National Institute of Standards and Technology of the USA (see \cite{nist}) are studying spectrum sharing. 
Spectrum sharing is also an important area in the efforts (focusing on 3.5 GHz band) of Wireless Innovation Forum and Citizens Broadband Radio Service in USA (see \cite{cbrs}). 
India too has published a set of guidelines on spectrum sharing (see \cite{deptindia}). 
\textcolor{black}{Despite the significant interest in spectrum sharing, there has been very little work exploiting spectrum sharing for real-time traffic, and none of them consider perceived video quality or QoE.}
We address this gap in this paper.

Bandwidth sharing essentially makes more resources available for video delivery. However, when allocating network resources, it is important to look beyond network-level metrics like average resource allocation (e.g., average number of OFDM resource blocks allotted to a client in a 5G network), and consider client-level metrics like Quality of Experience (QoE). 
QoE of a client captures the client's overall video experience (like in \cite{joseph2012jointly}). In this paper, we model QoE of a client as the average of perceived video quality of all packets of the client. Perceived video quality of a packet in turn is modelled as a client-specific function of network resources allotted for the delivery of the packet. 
Perceived video quality can be assessed by using metrics like Mean Opinion Score (MOS) \cite{cermak2011relationship}, Peak Signal-to-Noise Ratio (PSNR) \cite{van2008traffic}, Structural Similarity  Index Measure (SSIM) \cite{wang2004video}, etc.
This (client-specific) model allows us to capture how same amount of network resources (e.g., time slots, or an OFDM resource block of 5G New Radio) can translate to very different perceived video quality for clients depending on the video they are viewing. For instance, a client viewing a video with a lot of detail (or moving scenes) typically needs more network resources to achieve high perceived video quality (see \cite{paris1999zero}).
Additionally, this model allows us to capture how same amount of network resources can lead to much higher perceived video quality for a cell-center client compared to that for a cell-edge client.

\subsection{Main Contributions} \label{key contributions}

{Major contributions} of this work are summarized below:
\begin{enumerate}
    \item We develop a novel framework for real-time video delivery allowing bandwidth sharing between multiple operators, aimed at QoE maximization to its clients (see Section \ref{sectionsystemmodel}). Our framework specifically considers QoE as a function of perceived video quality of the clients.
    \item We propose a joint allocation and sharing policy, with strong theoretical optimality guarantees  (see Section \ref{sectionjointsharing}).
    \item We evaluate and provide insights into the nature of QoE gains from bandwidth sharing using extensive simulations (see Section \ref{sectionperformanceevaluation}). 
    \textcolor{black}{
    In particular, we show that the QoE gains are higher when there is an imbalance in the arrival rates and/or channel conditions for the operators.}
\end{enumerate}

\subsection{Organization of the paper} \label{organisation}
We discuss related work in Section \ref{section_related_work}.
System model is discussed in Section \ref{sectionsystemmodel}.
Section \ref{sectionjointsharing} presents our joint allocation and sharing policy and a related theoretical result.
Simulation results are in Section \ref{sectionperformanceevaluation}.
We conclude in Section \ref{sectionconclusion}.


\section{Related Work} 
\label{section_related_work}

\begin{figure}[htbp]
\centerline{\includegraphics[scale=0.26]{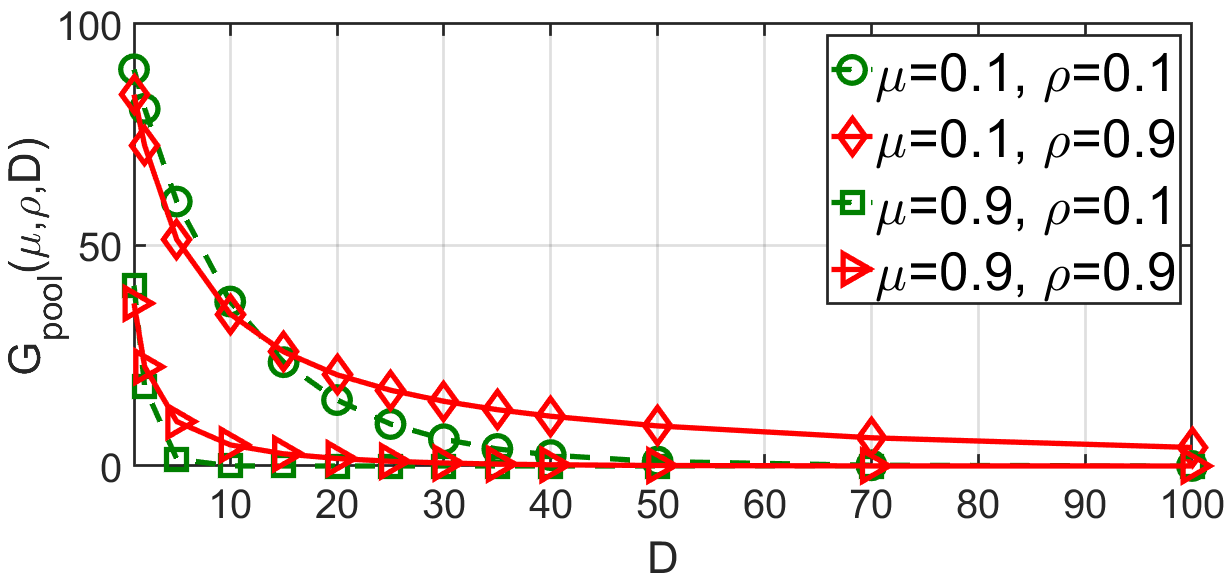}}
\caption{Improvement in success probability with sharing is higher for shorter deadlines ($D$)  }
\label{percent_improv_vs_D}
\end{figure}

\textbf{On real-time traffic related to queuing theory:} 
Queuing theory studies related to customer impatience (and reneging after lapse of deadline) provide useful insights about  serving of real-time traffic. 
By taking into account the M/M/c queue with independent exponentially distributed customer waiting times, Palm's groundbreaking study \cite{palm1953methods} examines queuing systems with impatient customers. 
\textcolor{black}{Only a few papers like \cite{bar,kok,cosmetatos1985approximate,boots1999multiserver}, however provide closed form expressions for useful metrics. 
Here, we take a closer look at the closed form expression in \cite{bar} for loss probability of an $M/M/1$ queue with deterministic impatience distribution.
An impatient customer with deadline $D$ is similar to a packet of real-time traffic (which too has a deadline). From \cite{bar}, the probability of successful service of a customer within deadline $D$ can be derived as:
\begin{equation}
    P_{{succ}}(\mu, \rho, D) =  \frac{1- e^{\mu D (\rho-1)}}{1-\rho e^{\mu D (\rho-1)}}.\nonumber
\end{equation}
Here $\mu$ is the average service rate of customers and $\rho$ is the traffic intensity. To assess the value of bandwidth sharing, we evaluate the percentage improvement in probability of successful service with sharing relative to that without sharing. 
Queue with sharing is obtained by pooling of arrivals and servers of two queues. Thus, it can be viewed as a new queue with double the arrival rate and service rate (i.e., $\mu$ is doubled whereas $\rho$ is unchanged). 
Hence, the percentage improvement in probability of successful service with sharing is
\begin{equation}
    G_{pool}(\mu,\rho,  D)  =\frac{P_{{succ}}(2\mu,\rho,  D) - P_{{succ}}(\mu,\rho, D)}{ P_{{succ}}(\mu, \rho, D)}.\nonumber
\end{equation}
Fig. \ref{percent_improv_vs_D}  shows how the gain drops as deadline $D$ increases. 
The larger pool of resources (resulting from sharing) is better suited to address real-time traffic's urgency in demand for resources.
Larger $D$ values can be viewed as traffic which is almost not real-time. A key takeaway here is that \textit{bandwidth sharing is especially attractive for real-time traffic} (more than for non-real-time traffic), in particular due to the urgency in its demand for bandwidth.}

Real-Time Queuing Theory (RTQT) introduced by Lehoczky addresses the limitations of using classic queuing theory (which typically focuses on average behaviour) for real-time systems. RTQT was explained using an $M/M/1$ queue with Earliest Deadline First (EDF) approach \cite{leho1}. More general settings were treated using RTQT in \cite{kruk}.

\textbf{On bandwidth/spectrum sharing between operators:} \cite{lil} proposed protocols for cellular networks to redistribute excess call traffic on a spectrum band, to spectrum bands with excess capacity. \cite{joshi,ikami2020dynamic, bennis2007inter} also propose spectrum sharing solutions and quantify the gains. These papers do not however consider real-time traffic.

\textbf{On scheduling of real-time traffic in wireless networks:}
There has been previous work on providing services for delay-constrained traffic in wireless networks. One of our key references is  \cite{hou2009theory}, which analyzed scheduling real-time traffic in unreliable wireless environments. \cite{hou2009theory} introduced a term \textit{timely throughput} to measure the amount of real-time traffic that is successfully delivered. Further, \cite{hou2013scheduling} and \cite{prk} developed scheduling policies with optimality guarantees for real-time traffic for various scenarios (e.g., rate adaptation, time-varying channels). Scheduling real-time traffic with hard
deadlines in a wireless ad hoc network by ensuring both timely throughput and data freshness guarantees for deadline-constrained traffic is considered in \cite{lu2018age}.
The system models in \cite{hou2009theory}, \cite{hou2013scheduling},\cite{prk} and \cite{lu2018age} rely on \textit{frame-based models} for arrival and scheduling of real-time traffic. In frame-based models (which is used in this paper too), all traffic arrives at the beginning of a frame and has to be served by the end of the frame. 
The approach in \cite{tsanikidis2021power} does not require this frame-based approach and uses an approach relying on randomization of the choice of transmitting links. Papers like \cite{age_optimal_sun}, \cite{timely_qi} have also approached this topic by formulating the problem in terms of \textit{age of information} and developed scheduling algorithms using deep reinforcement learning.
Note that these previous works do not however explore solutions leveraging bandwidth sharing.
\cite{ICC_paper} proposes a solution employing bandwidth sharing for serving real-time traffic, though it uses a simplified system model with a simplified resource allocation model and ignores perceived video quality.

\textbf{Optimizing QoE for wireless network}: \cite{huang_joint_source}, 
\cite{fu2010systematic}, 
and \cite{joseph2012jointly} propose network optimization to maximize metrics reflecting QoE or perceived video quality of users in a wireless setting. 
\cite{huang_joint_source} and \cite{fu2010systematic} use PSNR metric for perceived video quality whereas \cite{joseph2012jointly} considers a more general QoE metric modelled as a function of PSNR and SSIM values over time. However, the works too do not consider bandwidth sharing.

\section{System Model}\label{sectionsystemmodel}

\begin{table}[]
\caption{Notation}
    \centering
    \begin{tabular}{p{0.06\textwidth} p{0.38\textwidth}}
    \hline
\textbf{Notation}  & \textbf{Description} \\
\hline
\hline
 ${\mathcal O}$    &      Set of all operators\\
        \\[-1em]
 $i, j$      &    Indices used for operators\\
        \\[-1em]
${\mathcal R}$ &     Set of all regions
\\
        \\[-1em]
$r$     &       Index used for a region\\
        \\[-1em]
 ${\mathcal N}^{i}_{r}$ & Set of all clients in the $r^{th}$ region of the $i^{th}$ operator\\
  \\[-1em]
 ${\mathcal N}^{i}_{r}(k)$ & Subset of ${\mathcal N}^{i}_{r}$ comprising of clients with packet arrival in period $k$\\
        \\[-1em] 
        $n$ &   Index used for a client\\ 
        \\[-1em]
             $\mathcal{K}$ &  Set of all periods\\
        \\[-1em]
         $k$ &  Index of a period\\
        \\[-1em]
        $T$ & Number of timeslots within a period\\
        \\[-1em]
         $ A^{i}_{r,n}(k)$  & Number of packet arrivals of $n^{th}$ client in the $r^{th}$ region of the $i^{th}$ operator in the $k^{th}$ period\\
      \\[-1em]
       $ \alpha^{i}_{r}$  &    Average number of packet arrivals per period of each homogeneous client of the $i^{th}$ operator in the $r^{th}$ region \\
           \\[-1em]
       $ A^{i}_{r}(k)$  & Number of packet arrivals across all clients in the $r^{th}$ region of the $i^{th}$ operator in the $k^{th}$ period\\
       \\[-1em]
       $c_{r,n}^{i}$ & \textcolor{black}{Channel capacity} of the $n^{th}$ client in $r^{th}$ region of $i^{th}$ operator (number of bits that can be delivered to the client in one timeslot)\\
      \\[-1em]
     $\tau_{r,n}^{i}(k)$ & Number of timeslots alloted to  the $n^{th}$ client in $r^{th}$ region of $i^{th}$ operator in the $k^{th}$ period \\
       \\[-1em]
       $Q_{r,n}^{i}(.)$ & Perceived video quality of the $n^{th}$ client in $r^{th}$ region of $i^{th}$ operator \\
       \\[-1em]
       $Q_{min}$ & Minimum acceptable perceived video quality\\
       \\[-1em]
       $b_{r,n}^{i}(k)$ & Acceptable video quality indicator which takes value one when the perceived video quality $Q_{r,n}^{i}(k)$ is atleast $Q_{min}$, and is zero otherwise\\
       \\[-1em]
         $\delta_{r,n}^{i}(k)$  & Quality debt of the $n^{th}$ client in the $r^{th}$ region of the $i^{th}$ operator upto the $k^{th}$ period\\
         \\[-1em]
      $q_{r,n}^{i}$ & Minimum timely delivery rate requirement of the $n^{th}$ client in the $r^{th}$ region of the $i^{th}$ operator\\
       \\[-1em]
         ${S^{j\rightarrow i}_{r}(k)}$   & Number of timeslots of the $j^{th}$ operator shared with the $i^{th}$ operator in the $r^{th}$ region in the $k^{th}$ period\\
         \\[-1em]
         $\zeta^{(i,j)}$ & Bound on difference in sharing of any two operators\\
         \\[-1em]
         $\sigma^{i\rightarrow j}(k)$  & Sharing debt owed by operator $i$ to operator $j$ upto $k^{th}$ period\\
         \hline
         \hline
    \end{tabular}
    \label{tab:my_label}
\end{table}

We consider the downlink of a wireless system spanning a set of regions ${\mathcal R}$ served by a set of operators $\mathcal O$. 
Let ${\mathcal N}^{i}_{r}$ denote the set of clients in region $r$ of operator $i$.
Time is divided into timeslots and $T$ consecutive timeslots form a \textit{period}.

\textbf{Wireless channel:} 
Let $c_{r,n}^{i}$ denote the channel capacity of the $n^{th}$ client in the $r^{th}$ region of the $i^{th}$ operator, representing the number of bits that can be delivered to the client in one timeslot. 
This channel model allows capturing heterogeneous wireless channels, i.e., channels that can depend on client, operator and region. For instance, we can set a higher $c_{r,n}^{i}$ value for a client with good channel conditions (e.g., cell-center), when compared to another client with poor channel conditions (e.g., cell-edge). 
Further, operator-specific channel conditions can be set using channel capacity choices with $c_{r,n}^{i} \neq c_{r,n}^{j}$.

\textbf{Interference model:} 
We assume that an operator can transmit simultaneously in different regions without the transmissions interfering with each other.
There is no interference between transmissions of different operators in one region, and no interference across different operators in different regions, or same operator across different regions.

\textbf{Packet arrivals:} Like in \cite{prk}, we model video traffic of a client as a stream of packets with packet arrivals happening only at the start of a period. 
For client $n$ in region $r\in{\cal  R}$ of operator $i\in {\cal O}$, let  $A^{i}_{r,n}(k) \in \{0,1\}$ be the indicator function indicating whether packet for the client arrived in period $k$. 
We model $\left\{ A^{i}_{r,n}(k):k\ge 1\right\}$ as a stationary irreducible Markov process with finite state space and independent for any two clients. We have considered scenarios with correlated arrivals in simulations (see Section \ref{section simulation results}).
Also, let  $A^{i}_{r}(k)$ denote the total number of packet arrivals in period $k$ of all the clients in region $r\in{\cal  R}$ of operator $i\in {\cal O}$.


\textbf{Resource allocation}: A key consideration is whether typical packet sizes can be accommodated within a period. Given that 5G supports data rates ranging from 10 to 1000 megabits per second (Mbps), real-time video packets (typically between 400 and 1,500 bytes) can be efficiently transmitted within a 20 ms period. Let $\tau^{i}_{r,n}(k)$ denote the number of timeslots allotted to the $n^{th}$ client in the $r^{th}$ region of the $i^{th}$ operator in the $k^{th}$ period.
Thus, $\sum_{n\in{\mathcal N}^{i}_{r}}\tau^{i}_{r,n}(k) \le T$.
We assume that timeslots are allotted only to a client with a packet arrival in period $k$ (i.e., $\tau_{r,n}^{i}(k) =0$ if $ A_{r,n}^{i}(k)=0$). 
The number of bits delivered to the client is $c^{i}_{r,n}\tau^{i}_{r,n}(k)$.
Thus, the number of bits delivered to a client depends on the channel conditions and resource (timeslots) allocation.
For ease of exposition, we consider timeslots as the network resource. 
Our framework can be extended to cover systems using other types of \textit{orthogonal} resources, such as (OFDM) Resource Blocks in a 5G New Radio system.

\textbf{Quality of Experience:} We model  perceived video quality of the $n^{th}$ client in $r^{th}$ region of $i^{th}$ operator in the $k^{th}$ period as $Q^{i}_{r,n}(\tau_{r,n}^{i}(k))$. That is, it is a \textcolor{black}{twice-differentiable increasing concave} function of the number of timeslots allotted to the client.
Further, the client-dependent nature of the function $Q^{i}_{r,n}$ allows for same number of allotted timeslots to map to different perceived video quality for different clients. This allows for considering variations in the nature of real-time video delivered to different clients. 
The client-dependence of $Q^{i}_{r,n}$ also allows us to consider heterogeneity of clients' channels. For instance, $Q^{i}_{r,n}$ can depend on $c^{i}_{r,n}$ (see example in \eqref{simqualitymodel}).
This allows the same number of allotted timeslots to lead to higher perceived quality for a cell-center client compared to a cell-edge client. 

The perceived video quality across periods dictates the \textit{Quality of Experience (QoE)} of a client. We model the QoE of a client as the average of video quality perceived by the client across all periods.


\textbf{Timely delivery requirement:} A packet that arrives at the start of a period must be transmitted within the period (i.e., within $T$ timeslots) for its \textit{timely} (or successful) delivery, or is dropped otherwise. \textcolor{black}{Some examples of $T$ for real-world applications are given in Table \ref{latency}.}
Further, the packet needs to be transmitted within the period with an acceptable quality of at least $Q_{min}$. $Q_{min}$ can be set by the network operator to reflect minimum acceptable video quality (the minimum can be made client-specific, though we avoid it to keep the notation simple).
\textcolor{black}{Let $b_{r,n}^{i}(k)$ be the acceptable video quality indicator for the packet that arrives in period $k$ of the $n^{th}$ client in $r^{th}$ region of $i^{th}$ operator.}
$b_{r,n}^{i}(k)$ takes value of one when the  packet's perceived video quality $Q^{i}_{r,n}(\tau^{i}_{r,n}(k))$ is greater than or equal to $Q_{min}$, and is otherwise zero.
Also, let $b_{r,n}^{i}(k)=0$ if no packet arrives for the client in period $k$.

A key requirement that we consider is related to the fraction of packets that are delivered to a client in a timely manner  with acceptable quality, referred to as \textit{\textcolor{black}{timely delivery rate}} of the client.  
Client $n\in{\mathcal N}^{i}_{r}$ requires timely delivery rate of at least $q_{r,n}^{i}$, which is expressed as the following probabilistic requirement for a small positive constant $\xi_1$:
\begin{align}
    &&\text{Prob} \left\{\frac{1}{K}\sum_{k=1}^{K} b_{r,n}^{i}(k) \ge  q_{r,n}^{i} - \xi_1 \right\} \rightarrow \text{1},\hspace{1cm} \nonumber \\ &&\text{~as~} K\rightarrow\infty , \ \forall \ {n} \in {\mathcal N}^{i}_{r}, \ \forall \ {i} \in {\mathcal O},\ \forall \ {r} \in {\mathcal R}. \label{throughput constraint} 
\end{align}
For example, consider the acceptable packet loss of 5\% mentioned in \cite{webex}, which can be captured by setting $q_{r,n}^{i} $ as 0.95 times the packet arrival rate for the client.
Timely delivery rate metric is an extension of the timely throughput metric in \cite{prk}. 

\vspace{-0.09cm}
\subsection{Bandwidth Sharing Model} 
In any region, an operator can share its resources (timeslots in a period) with another operator. 
Note that this sharing is region-specific, and different sharing is possible in different regions. 
Let ${S^{j\rightarrow i}_{r}(k)}$ denote the number of timeslots of the $j^{th}$ operator shared with the $i^{th}$ operator in the $r^{th}$ region in the $k^{th}$ period. 
Here, ${S^{i\rightarrow i}_{r}(k)}$ is the number of timeslots of the $i^{th}$ operator used for its own clients in $r^{th}$ region in the $k^{th}$ period.

Following is a key constraint in our sharing framework which bounds the difference in average number of timeslots shared by any pair of operators ${i,j} \in {\mathcal O}$, \textcolor{black}{across all regions}:
\begin{align}
 \text{Prob} \left\{ \hspace{-0.1cm} \frac{1}{K} \left | \sum_{k=1}^{K} \sum_{r \in \mathcal {R}} \hspace{-0.05cm} S^{j\rightarrow i}_{r}(k)  - \hspace{-0.1cm}\sum_{k=1}^{K} \sum_{r \in \mathcal {R}} S^{i\rightarrow j}_{r}(k)\right |\hspace{-0.05cm} \leq \zeta^{(i,j)} \hspace{-0.1cm} + \hspace{-0.05cm}\xi_2 \right\} \nonumber \\  \rightarrow \text{1},   \text{~as~} K\rightarrow\infty, \label{sharing constraint} 
\end{align}
where $\xi_2$ is a small positive constant. 
Here the parameter $\zeta^{(i,j)}$ controls the relative amount of sharing between operators $i$ and $j$.
In particular, for small $\zeta^{(i,j)}$, \eqref{sharing constraint} essentially ensures that each operator gives roughly as much it gets from another operator as part of sharing \textcolor{black}{(across all regions)}.  
Thus \eqref{sharing constraint} incentivizes sharing (an operator receives only as much as it gives) and also limits over-sharing (an operator gives only as much as it receives). \textcolor{black}{
Note that more spectrum is used with bandwidth sharing by each operator, and this can lead to additional interference in practice for some regions. Region-specific weights in \eqref{sharing constraint} can discourage sharing for such regions.}

Without sharing, the number of timeslots that can be allotted by operator $i$ to its client in region $r$ is bounded by ${S^{i\rightarrow i}_{r}(k)}\leq T$.
With sharing, the number of timeslots that can be allotted by operator $i$ to its clients in region $r$  is bounded by the sum of ${S^{i\rightarrow i}_{r}(k)}$ (timeslots owned by the operator) and
$\sum_{j\in {\mathcal O}\setminus \{i\}}S^{j\rightarrow i}_{r}(k)$ (timeslots shared by other operators), i.e., we have:
 \begin{align}
 \sum_{n\in {\mathcal N}^{i}_{r}}\tau_{r,n}^{i}(k) & \leq \sum_{j\in {\mathcal O}}S^{j\rightarrow i}_{r}(k), \ \forall {i} \in {\mathcal O}, \forall{r} \in {\mathcal R}, \forall{k} \in {\mathcal K}  \label{maxschedule} 
 \end{align}
 
A few additional constraints on sharing are given below:
\begin{align}
&  \qquad \quad S^{j\rightarrow i}_{r}(k) \geq 0,  \qquad \forall {i,j} \in {\mathcal O}, \forall{r} \in {\mathcal R}, \forall{k} \in {\mathcal K};\label{nonneg}\\
& \qquad \sum_{i\in {\mathcal O}}S^{j\rightarrow i}_{r}(k) \leq T, \qquad  \forall {j} \in {\mathcal O}, \forall{r} \in {\mathcal R}, \forall{k} \in {\mathcal K};  \label{maxshare} 
\end{align}


\eqref{nonneg} requires that ${S^{j\rightarrow i}_{r}(k)}$ are non-negative.
\eqref{maxshare} captures that maximum number of timeslots available to an operator for its own use and for sharing with other operators is $T$. 

\subsection{Asymptotically Optimal Policy} \label{feasibleregion}

Observe from the preceding discussion that the key decision variables involved are the following:
\begin{itemize}
    \item how many timeslots to be allotted to each client, i.e., $\forall k$ deciding
    $\bm{\tau}(k) = \left( \tau_{r,n}^{i}(k): \ \forall{n} \in {\mathcal N}^{i}_{r}, \ \forall{r} \in {\mathcal R}, \  \forall {i} \in {\mathcal O} \right) $, and
    \item how much to share, i.e., $\forall k$ deciding $\mathbf{S}(k) = \left(S^{j\rightarrow i}_{r}(k): \ \forall{r} \in {\mathcal R}, \  \forall {i,j} \in {\mathcal O}\right)$. 
\end{itemize}
 A \textit{policy} specifies the timeslot allotment $\bm{\tau}(k)$ and sharing  variables $\mathbf{S}(k)$ for each period $k$, and has to satisfy constraints \eqref{maxschedule}-\eqref{maxshare}.

 We let $\bm{q} = \left( q_{r,n}^{i}: n\in {\mathcal N}^{i}_{r}, \ \forall{r} \in {\mathcal R}, \  \forall {i} \in {\mathcal O}  \right)$ and $\bm{\zeta} = \left( \zeta^{(i,j)}: {i,j} \in \mathcal O\right)$.
   Timely delivery rate requirement $\bm{q}$ and sharing bound $\bm{\zeta}$ is said to be \textit{feasible}, if there exists a policy satisfying
\eqref{throughput constraint}-\eqref{maxshare}.  
Timely delivery rate requirement $\bm{q}$ and sharing bound $\bm{\zeta}$ is said to be \textit{strictly feasible}, if there exists a constant $\kappa$ with $0<\kappa<1$, such that modified timely delivery rate requirement $\bm{q}/\kappa$ and sharing bound $\kappa\bm{\zeta}$ is feasible. 

A policy that maximizes the following objective function
\begin{align}
  \lim_{K \to \infty} \frac{1}{K} \sum_{k=1}^{K} \sum_{r \in \mathcal {R}} \sum_{i\in \mathcal O} \sum_{{n} \in {\mathcal N}^{i}_{r}} Q^{i}_{r,n}(\tau_{r,n}^{i}(k))
  \label{objfunctionexpression}
\end{align}
subject to constraints \eqref{throughput constraint}-\eqref{maxshare} is said to be an \textit{asymptotically optimal} policy. Such a policy provides close-to-optimal performance when we consider large number of periods $K$.

\section{Online Joint Allocation and Sharing Policy} \label{sectionjointsharing}


Here, we present our joint allocation and sharing policy, and discuss a  related optimality result (Theorem \ref{theorem_feasibility_optimal}).
Designing an "optimal" policy is not straightforward, especially due to the time averaging involved in the timely delivery rate constraint (\ref{throughput constraint}) and sharing constraint (\ref{sharing constraint}). To tackle this, we utilize `debt parameters' or virtual queues $\delta_{r,n}^{*i}(k)$ and $\sigma^{*i\rightarrow j}(k)$, similar to those in \cite{prk, neely} etc.
Here $\delta_{r,n}^{*i}(k)$ is the virtual queue tracking \textit{quality debt} of the $n^{th}$ client in the $r^{th}$ region of the $i^{th}$ operator  up to the $k^{th}$ period (similar to that in \cite{prk}).
$\sigma^{*i\rightarrow j}(k)$ denotes the \textit{sharing debt} owed by operator $i$ to operator $j$ up to the $k^{th}$ period. 
Note that we are distinguishing variables related to our policy using superscript ${}^*$.

Our online policy maximizes the following function in each period $k$:
\begin{IEEEeqnarray}{rc}
\hspace{-0.5cm} & f(\bm{\tau}(k), \mathbf{S}(k),\bm{\delta}(k),\bm{\sigma}(k)) = \nonumber\\
& \sum_{r \in \mathcal {R}} \sum_{i\in \mathcal O} \sum_{{n} \in {\mathcal N}^{i}_{r}}\hspace{-0.1cm}\delta_{r,n}^{i}(k)b_{r,n}^{i}(k) + V \sum_{r \in \mathcal {R}} \sum_{i\in \mathcal O}\sum_{{n} \in {\mathcal N}^{i}_{r}} Q^{i}_{r,n} \left(\tau^{i}_{r,n}(k) \right)   \nonumber\\ 
   & - \sum_{i\in \mathcal O}\sum_{j\in {\mathcal O}\setminus\{i\}}\sigma^{i\rightarrow j}(k)\left (\sum_{r \in \mathcal {R}} S^{j\rightarrow i}_{r}(k)-\sum_{r \in \mathcal {R}} S^{i\rightarrow j}_{r}(k)\right),
   \label{eqn:fdefn}
\end{IEEEeqnarray}
where $\bm{\delta}(k)$ is a multi-dimensional array with entries $\delta_{r,n}^{i}(k)$ for each $n\in {\mathcal N}^{i}_{r}, \ \forall{r} \in {\mathcal R}, \  \forall {i} \in {\mathcal O} $, 
and $\bm{\sigma}(k)$ is a multi-dimensional array with entries $\sigma^{i\rightarrow j}(k)$ for each ${i,j} \in \mathcal O$. 
$V$ is a positive constant that affects the performance of the policy.
Our policy is given below: 
	\\\line(1,0){251.0}\vspace{-.3cm}
	\begin{center}
	\textbf{Online Joint Allocation and Sharing Policy}  \label{sharing scheduling policy}\vspace{-.5cm}
	\end{center} 
	\line(1,0){251.0} 
	\\Timeslots allotted $\bm{\tau}^*(k)$ and sharing $\mathbf{S}^*(k)$ for period $k$ are determined by solving the following optimization problem:
	   \begin{subequations}
\begin{align}
(\bm{\tau}^*(k), \mathbf{S}^*(k)) = 
& \underset{  (\bm{\tau}, \mathbf{S}) }{\text{argmax}}
& & \hspace{-0.1cm}f(\bm{\tau}, \mathbf{S},\bm{\delta}^*(k),\bm{\sigma}^*(k)) \label{optmzn_9a} \\
\
& \qquad  \text{s.t.} & &  \eqref{maxschedule}, \eqref{nonneg}  \mbox{ and }  \eqref{maxshare}. \label{optmzn_9b} 
    \end{align}
\end{subequations}
$\delta_{r,n}^{*i}(0) =\sigma^{*i\rightarrow j}(0)=0$, and are updated in each period $k>0$ as follows:
\begin{align}
  \delta_{r,n}^{*i}(k+1) & =
  \begin{cases}
      \max( \delta_{r,n}^{*i}(k)+q_{r,n}^{i}-b_{r,n}^{*i}(k), 0), \\
      \hspace{1.9cm} \forall n \in \mathcal{N}^{i}_{r}(k), \ i\in \mathcal{O}, \ r \in \mathcal{R} \\
      \delta_{r,n}^{*i}(k), \\
      \hspace{1.3cm} \forall n \in \mathcal{N}^{i}_{r} \setminus \mathcal{N}^{i}_{r}(k), \ i\in \mathcal{O}, \ r \in \mathcal{R}
  \end{cases} 
 \label{debts1} 
 \\ \sigma^{*i\rightarrow j}(k+1) & = \max( \sigma^{*i\rightarrow j}(k)+  \label{debts2} \\ \nonumber
   & \qquad \sum_{r \in \mathcal {R}} S^{*,j\rightarrow i}_{r}(k)-\sum_{r \in \mathcal {R}} S^{*,i\rightarrow j}_{r}(k)-\zeta^{(i,j)}, 0) ,  \\ \nonumber & \qquad \qquad \forall {i \in \mathcal O}, \ {j\in \mathcal O}\setminus\{i\}, \ { k \in \mathcal K}.  \vspace{-.25cm}
\end{align} 
\line(1,0){251.0} 

Observe that the above policy is an \textit{online} policy making decisions solely based on information readily available in period $k$. It  does not explicitly include long term time-average requirements \eqref{throughput constraint}  and \eqref{sharing constraint} as constraints. 
Rather, they are met by ensuring that the debts tracked using $\delta_{r,n}^{*i}(k)$ and $\sigma^{*i\rightarrow j}(k)$ are not too high.
In particular, a client with low number of timely packets deliveries with acceptable quality, will tend to have a high value of $\delta_{r,n}^{*i}(k)$ (see \eqref{debts1}). This high value leads to a higher chance of providing higher perceived quality to that client, since $\delta_{r,n}^{*i}(k)$ scales $b_{r,n}^{*i}(k)$ in the objective function \eqref{eqn:fdefn}. 
Similarly, an operator $i$ that has not shared much will tend to have a high value of  $\sigma^{*i\rightarrow j}(k)$ (see \eqref{debts2}), and this leads to higher chance of sharing by the operator in subsequent periods, since $\sigma^{*i\rightarrow j}(k)$ scales $S^{*,i\rightarrow j}_{r}(k)$ in the objective function \eqref{eqn:fdefn}. 


 \subsection{Optimality} \label{optimality}

Following theorem provides a strong  theoretical performance guarantee for our joint allocation and sharing policy. Its proof is in Appendix \ref{sectionappendix}. 

 \begin{theorem}
 \label{theorem_feasibility_optimal}
 
  For any strictly feasible timely delivery rate requirement $\bm{q}$  and sharing bound $\bm{\zeta}$, the allocation and sharing policy $\left(\left(\bm{\tau}^*(k), \mathbf{S}^*(k)\right):k\ge 1 \right)$ 
  satisfies \eqref{throughput constraint}-\eqref{maxshare} and achieves value for the objective function  \eqref{objfunctionexpression} that is within $\mathcal{O}\left (1/V \right )$  that of asymptotically optimal policy.
  \end{theorem}
Thus, our policy achieves feasibility whenever possible, and attains \textcolor{black}{total QoE (summed over all clients)} close to that of the optimal (asymptotic) policy.



 \section{Performance evaluation} \label{sectionperformanceevaluation}
In this section, we present simulation results for our joint allocation and sharing policy. We primarily assess benefits of sharing by evaluating \textit{percentage improvement in total QoE of all clients} with sharing compared to that without sharing. 

\subsection{Simulation Settings}\label{subsectionsimulationsettings}
We consider two operators serving two regions (i.e., ${\cal R}=\{1,2\}$) with 30 clients each. 
To ensure roughly equal sharing between operators, we set $\zeta^{(i,j)}$ as a small value of 0.001. For simplicity, we assume all the clients of an operator in a region have the same average packet arrival per period.
Let $\alpha^i_{r}$ denote the average number of packet arrivals per period for each client in region $r\in{\cal  R}$ of operator $i\in {\cal O}$. We set $\alpha^1_{1}$ = $\alpha^2_{2}$ and $\alpha^1_{2}$ = $\alpha^2_{1}$, so that an operator will have a relatively low arrival rate in one region and a high arrival rate in the other \textcolor{black}{(performance improvements may be limited without such an imbalance in arrival rates due to \eqref{sharing constraint})}.  Throughout the simulation section, we model the packet arrival process for each client using a Bernoulli distribution.
We set $q_{r,n}^{i} = 0.95$ times the packet arrival rate for each client, based on typical values like 5\% for allowable packet loss (see \cite{webex}). 
We set $c_{r,n}^{i}=10 \times 10^6$ bits per timeslot and $10000$ periods.


\subsubsection{Setting length of a period}
Inorder to set the length of a period, two parameters are taken into consideration - acceptable latency and typical interarrival time between video frames. The acceptable latency for various real-time applications are tabulated in \ref{latency}. Since our work deals with various real-time applications, we can set the allowable latency as 20 ms. Considering frame rate, the best frame rate for 4K videos is 50 or 60 fps depending on different geographical locations \cite{minitool}, \cite{DVDFab}. The interarrival time between frames can therefore be roughly approximated as 20 ms. Based on the acceptable latency and the interarrival time between frames, we set length of a period $T=20$ (the deadline of each packet too is twenty timeslots).

\begin{table}[]
\centering
\caption{Acceptable latency in various real-time applications}
\label{latency}
\begin{tabular}{|l|l|}
\hline
\textbf{Real time application} & \textbf{Acceptable latency (ms)} \\ \hline
Virtual Reality                & \textless 20   \cite{hou2017wireless}, \cite{hazarika2023towards}                  \\ \hline
Video conferencing             & \textless 150    \cite{itu911}                \\ \hline
Online gaming                  & \textless 50  \cite{centurylink}, \cite{screenbeam}, \cite{wiredshopper}                   \\ \hline
Live streaming                 & \textless 100  \cite{haivision}                  \\ \hline
\end{tabular}
\end{table}

 \subsubsection{Tau-Quality relation}
The most commonly used video codec is H.264 \cite{adobe}, \cite{wiki}, \cite{wowza}, \cite{microsoft}, \cite{zoom}, \cite{webex_VC}. Considering the bandwidth to be 10 MHz, the spectral efficiency to be 5 bps/Hz (based on typical values \cite{5GNR}, \cite{coleago}) and based on \cite{cermak2011relationship}, we obtained the following model for perceived video quality (unitless) for  H.264 video codec.
\begin{align}
   Q_{r,n}^{i}(\tau_{r,n}^{i})  =\frac{1}{0.8} \ln \left(\frac{\frac{\tau_{r,n}^{i}c_{r,n}^{i}}{T}+0.1}{0.4}\right)
   \label{simqualitymodel}
\end{align}

This choice of $ Q_{r,n}^{i}(\tau_{r,n}^{i})$ is increasing and concave in $\tau_{r,n}^{i}$. Further, it is increasing in $c_{r,n}^{i}$. In particular, a cell-edge client (with low $c_{r,n}^{i}$) will realize lower perceived video quality than a cell-center client. \textcolor{black}{The improvement in quality with each additional $\tau_{r,n}^{i}$ can be ascribed to the derivative of \eqref{simqualitymodel}. This derivative, analogous to the concept of marginal utility in economics, can be termed as marginal quality. }\\

360p is the lowest standard definition widely used for live streaming, offering a clear viewing experience without noticeable blur on smaller screens such as smartphones, tablets, and older computers or TVs \cite{boris},\cite{resi}.  A bitrate of around 0.4 Mbps is necessary for video with a minimum resolution of 360p \cite{minBRgoogle}, \cite{dacast}. Therefore, we chose 0.3 as the minimum acceptable perceived quality, or $Q_{min}$.

We refer to $\dfrac{dQ_{r,n}^{i}}{d\tau_{r,n}^{i}}$ as  marginal quality, and it  captures the improvement in QoE of a client for every additional timeslot unit allotted to the client. Note that 
\begin{align}
   \dfrac{dQ_{r,n}^{i}}{d\tau_{r,n}^{i}}  =\frac{1}{\gamma_Q} \left(\frac{0.4}{\tau_{r,n}^{i}+\frac{0.1T}{c_{r,n}^{i}}}\right). 
   \label{derivative}
\end{align}

\subsection{Simulation Results} \label{section simulation results}
\begin{figure}
\label{arrival_imbalance} 
\centering
\begin{subfigure}[b]{0.55\textwidth}
   \includegraphics[width=0.9\linewidth]{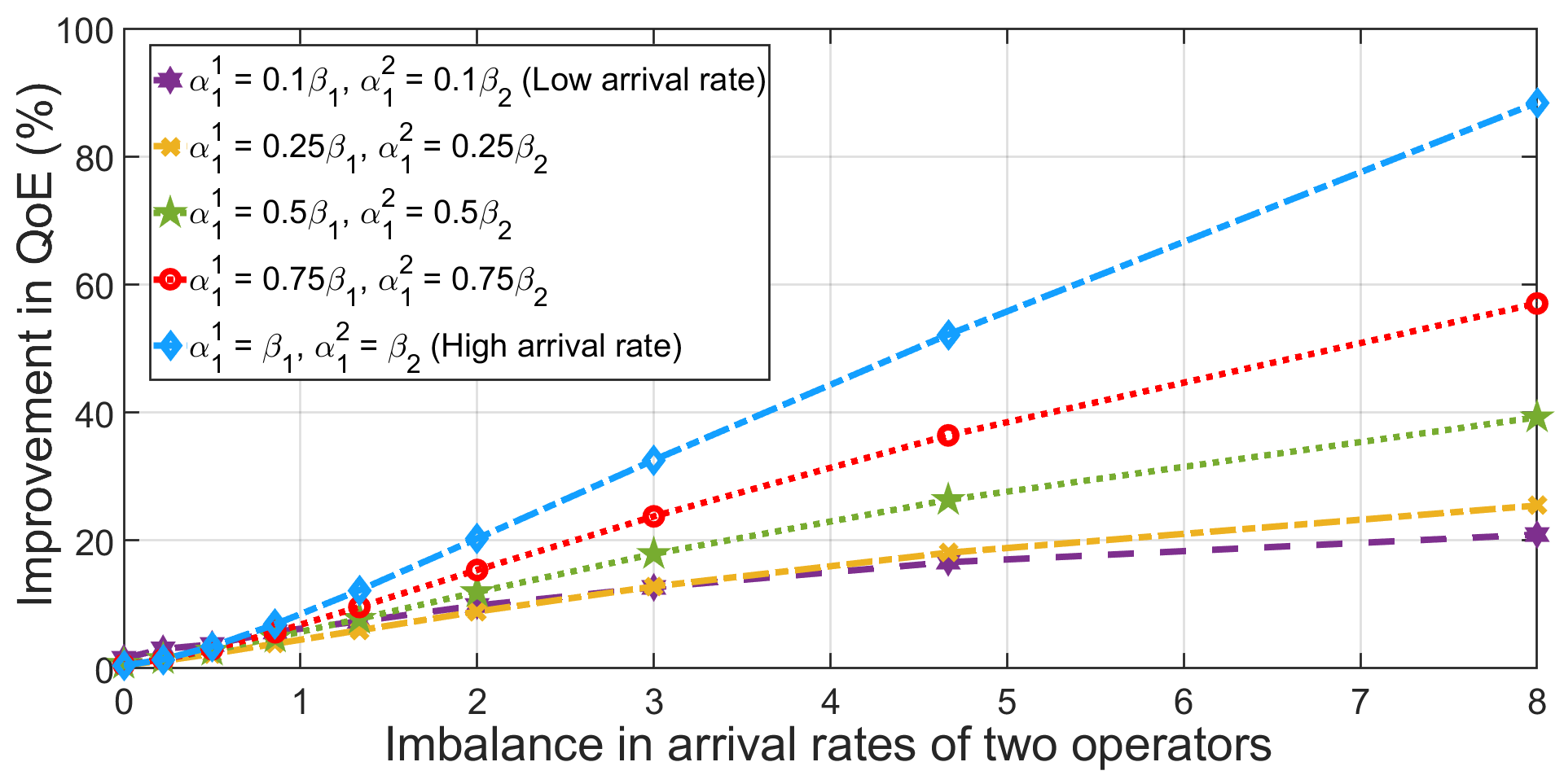}
   \caption{}
   \label{improv_QoE} 
\end{subfigure}
\begin{subfigure}[b]{0.55\textwidth}
   \includegraphics[width=0.9\linewidth]{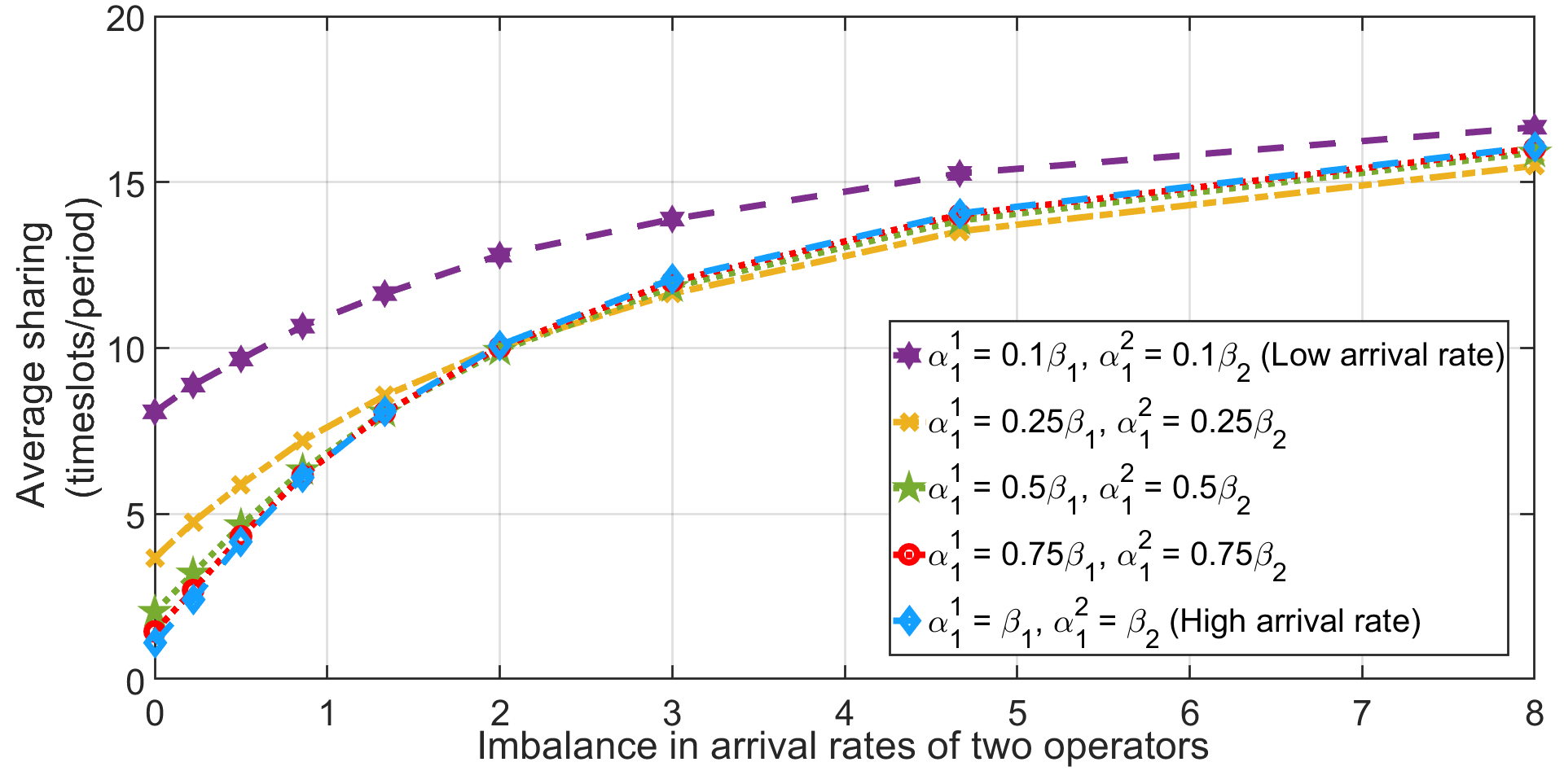}
   \caption{}
   \label{sharing}
\end{subfigure}
\begin{subfigure}[b]{0.55\textwidth}
   \includegraphics[width=0.9\linewidth]{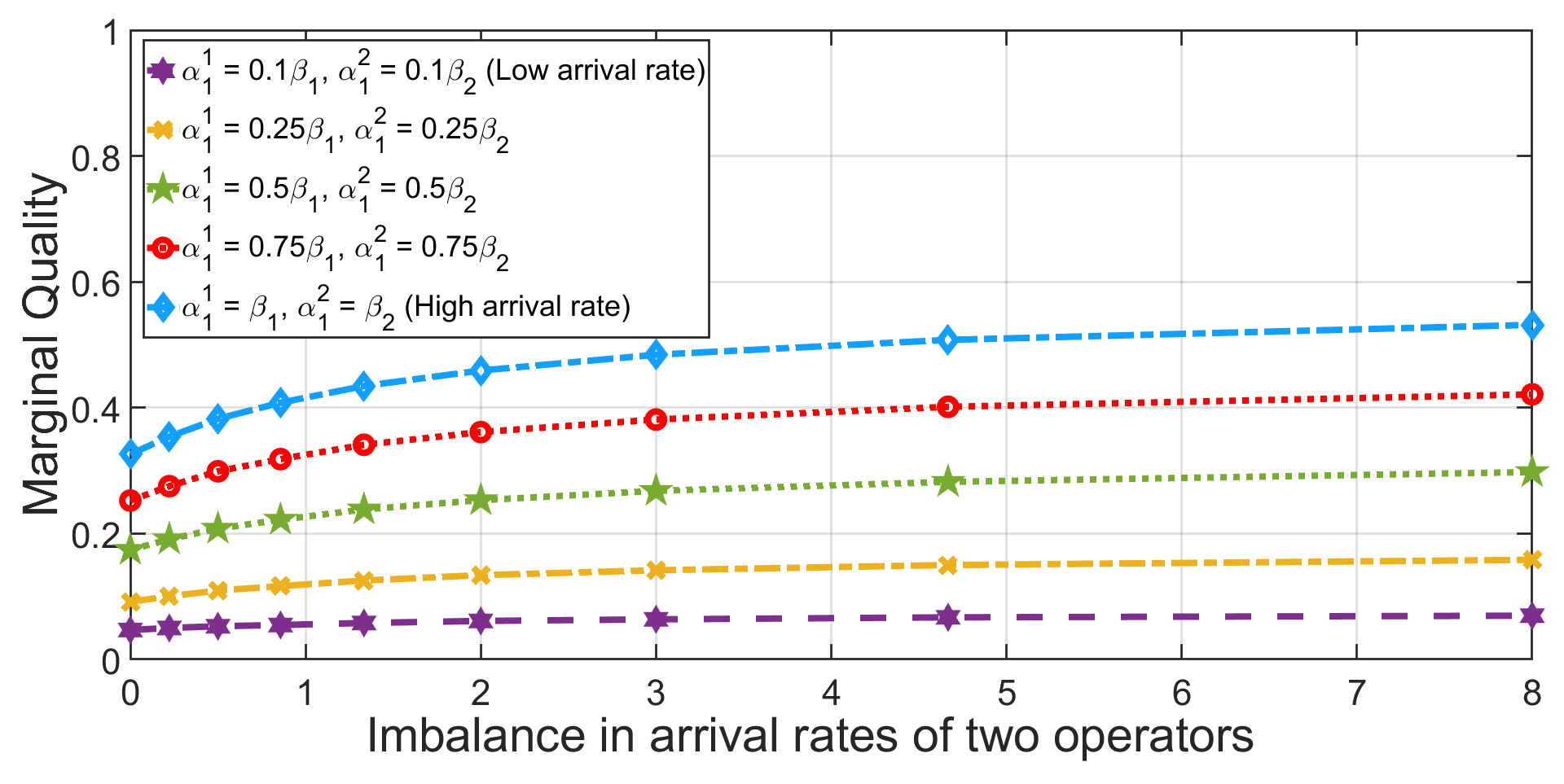}
   \caption{}
   \label{dQ}
\end{subfigure}
\caption[Imbalance in arrival rates]{Impact of imbalance in arrival rates between two operators: (a) Percentage improvement in QoE, (b) Average number of timeslots shared per period, (c) Marginal quality}\end{figure}

In Fig. \ref{improv_QoE}, we analyse the impact of imbalance in arrival rates which is measured as $\left( \left({\alpha^2_1}/{\alpha^1_1}\right)-1\right)$, \textcolor{black}{under homogeneous channel conditions ($c_{r,n}^{i}=10 \times 10^6$ bits per timeslot)}. Here we set arrival rates in region 1 and 2 as scaled versions of $\beta_1$ and $\beta_2 = 1- \beta_1$ respectively, where $\beta_1 \in \{0,0.05,0.1,...,0.5\}$. Here, the results indicate that \textit{higher the imbalance in arrival rates, higher are the improvements in QoE} with sharing. This is expected as unbalanced arrival rates allow more sharing, as the operator with low arrival rate in a region can give its resources to the other operator in that region, and the other operator reciprocates in the other region. 

\textcolor{black}{The gains in quality can be attributed to two key factors: average sharing and marginal quality.} \textcolor{black}{For low arrival rate imbalances,} the primary determinant of QoE improvement is average sharing (see fig. \ref{sharing}). \textcolor{black}{Whereas for higher arrival rate imbalances,} the average sharing remains relatively constant across various arrival rate patterns \textcolor{black}{and the marginal quality} (see fig. \ref{dQ}) becomes the pivotal factor influencing QoE improvement.

\begin{figure}[htbp]
\centerline{\includegraphics[scale=0.17]{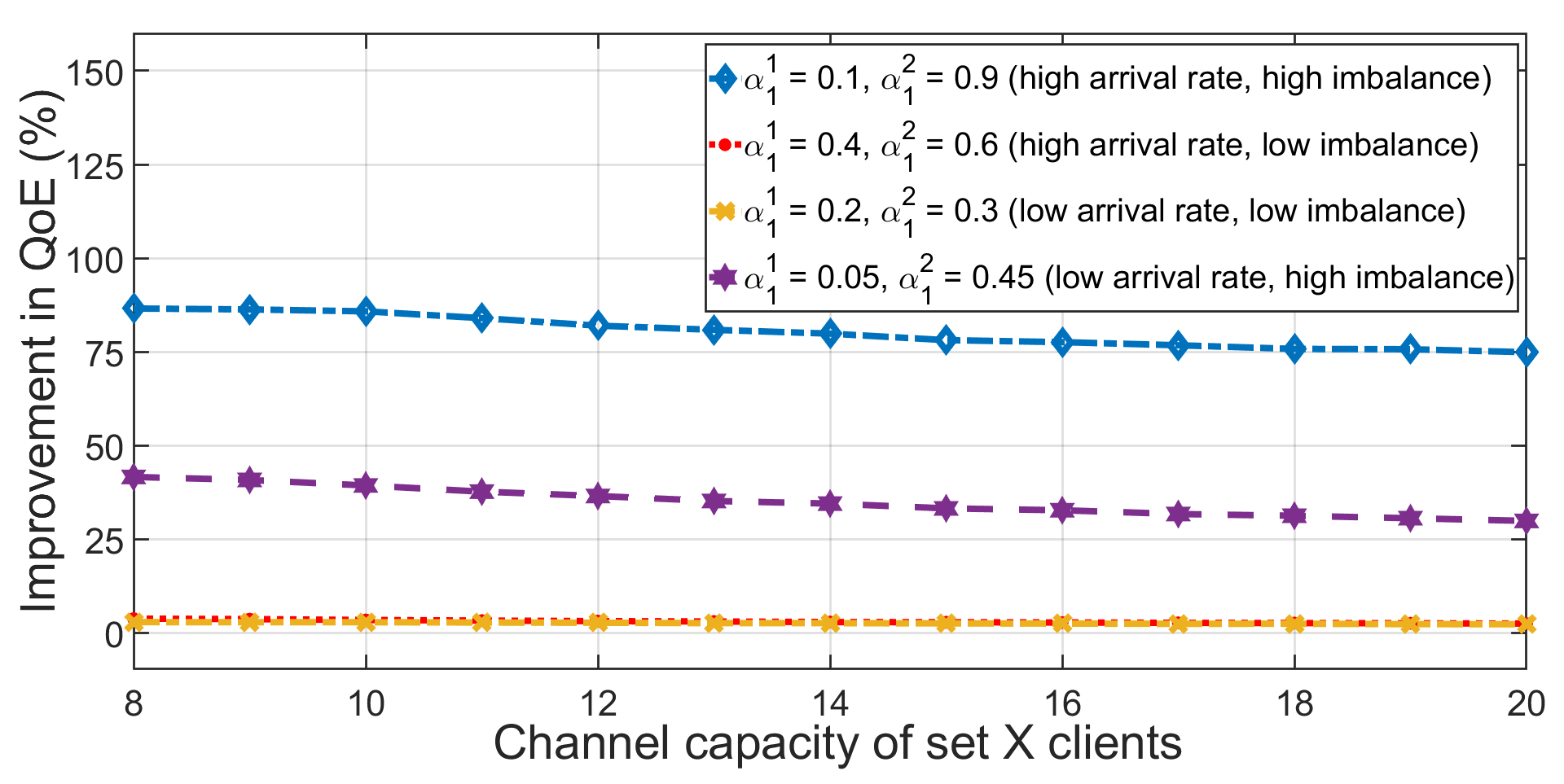}}
\caption{Percentage improvement in QoE with variation in channel capacity.
}
\label{vary_hetero}
\end{figure}

Using Fig. \ref{vary_hetero}, we explore the impact of  heterogeneous channel capacities on the QoE improvements.
Here, half of the clients (referred to as set Y) of an operator have channel capacity $10 \times 10^6$ bits per timeslot, and  all the remaining clients (referred to as set X) have equal channel capacity which is set to a value between $8 \times 10^6$ to $20 \times 10^6$ bits per timeslot. 
The results show that \textit{sharing is beneficial even for clients with less favorable channel conditions, and more generally under a wide range of channel conditions}. 
Further, the "flatness" of Fig. \ref{vary_hetero} indicates that the channel conditions do not have a significant impact on the improvement in QoE. 
This is expected since marginal quality is not impacted much by $c_{r,n}^{i}$ (see \eqref{derivative}).

\begin{figure}[htbp]
\centerline{\includegraphics[scale=0.17]{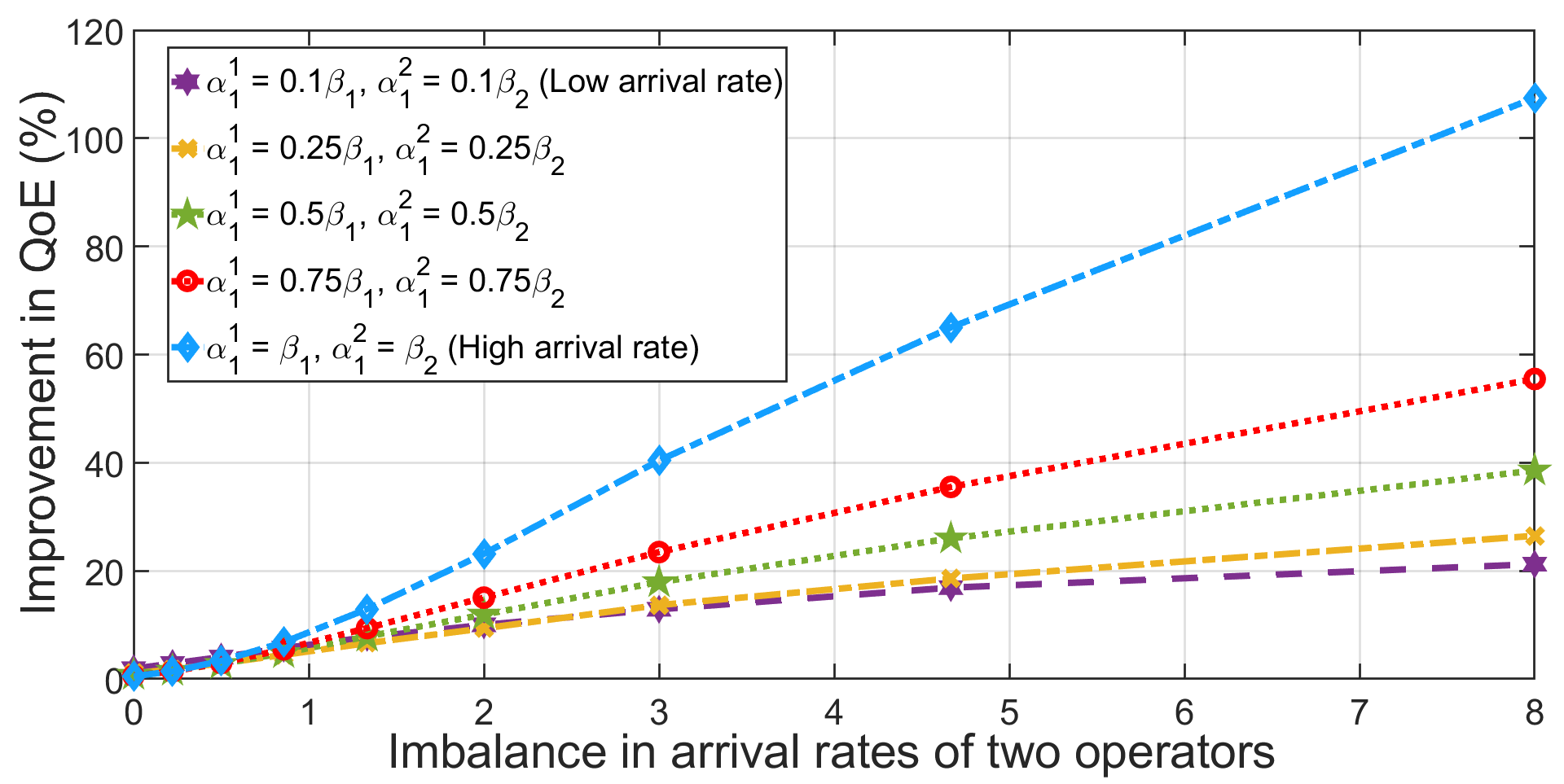}}
\caption{Impact of heterogeneous quality scaling factors}
\label{quality_scaling}
\end{figure}

In Fig. \ref{quality_scaling}, we delve into how QoE improvement varies based on the level of detailing in the videos streamed by clients. Unlike previous plots where all clients of both operators viewed similar content, here we assume that each operator caters to clients viewing highly detailed, moderately detailed, and less intricate videos. The results clearly demonstrate that \textit{in this practical setting also our joint allocation and sharing policy is able to achieve significant gains in QoE improvement}. This significant increase in QoE improvement can be attributed to the greater resource demand for clients engaged with more intricate content compared to those watching less detailed material. Consequently, the available resources might not suffice to ensure good QoE for these clients, resulting in lower QoE levels. Resource sharing helps to tremendously increase the already low QoE for high arrival rates. Conversely, when arrival rates are very low, clients already enjoy satisfactory QoE without sharing, leading to comparatively smaller QoE improvements.


\begin{figure}[htbp]
\centerline{\includegraphics[scale=0.17]{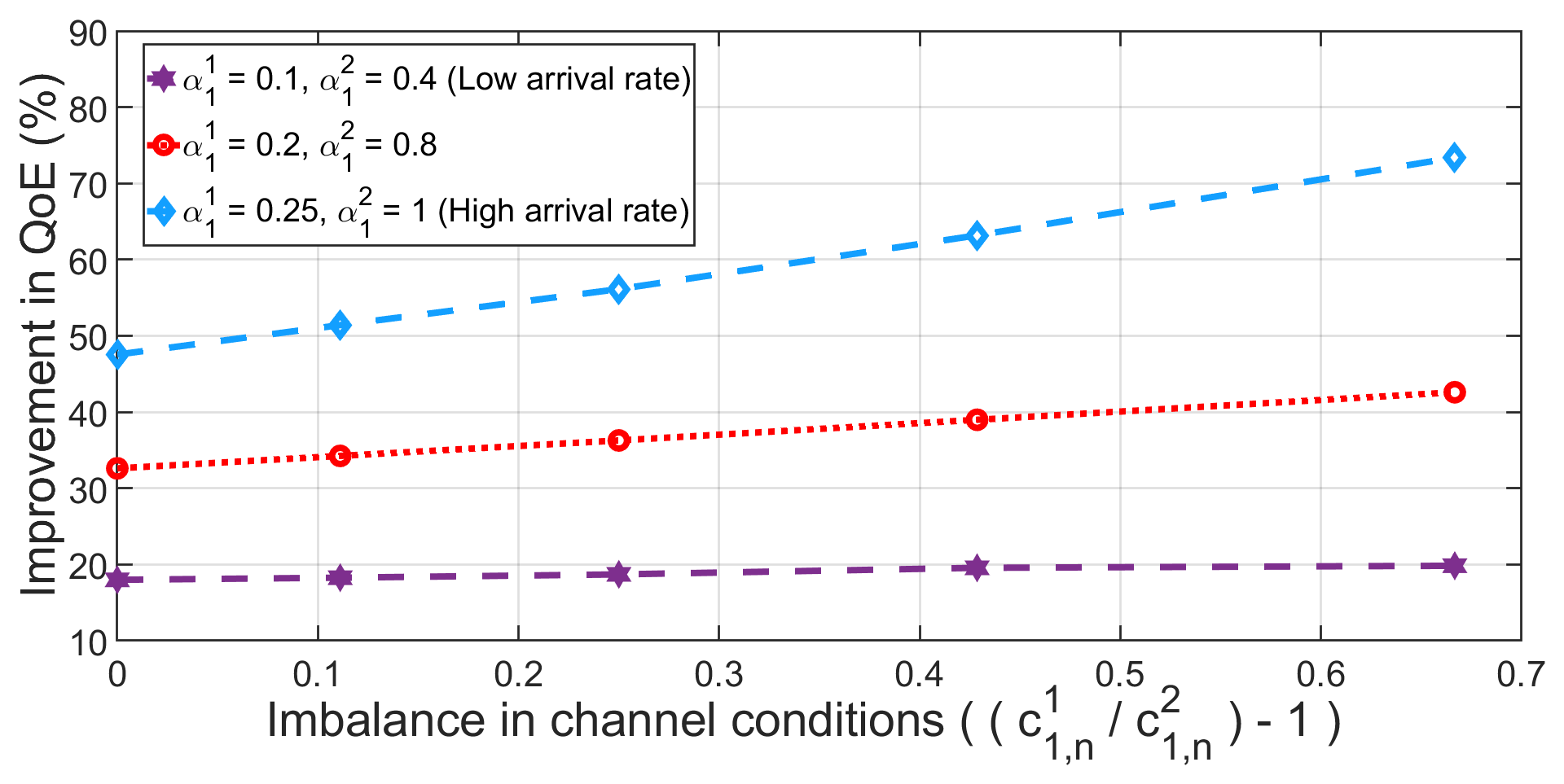}}
\caption{ Impact of coverage imbalance on QoE improvement}
\label{coverage}
\end{figure}

In Fig. \ref{coverage}, we study the impact of imbalance in coverage which is measured as $\left( \left({c_{r,n}^{1}}/{c_{r,n}^{2}}\right)-1\right)$. Here,  we fix the channel capacity of all clients of an operator in a region as $c_{r,n}^{1}$ = $10 \times 10^6$ bits per timeslot and vary the channel capacity of all clients of the other operator in the same region from $10 \times 10^6$ to $6 \times 10^6$ bits per timeslot. The trend shows that as the imbalance in coverage grows, there is a noticeable rise in QoE improvement due to an increase in average sharing. Even when there is low imbalance in the arrival rates between operators, \textit{a significant improvement in QoE is possible when operators have clients with different channel conditions}. 

\begin{figure}[htbp]
\centerline{\includegraphics[scale=0.17]{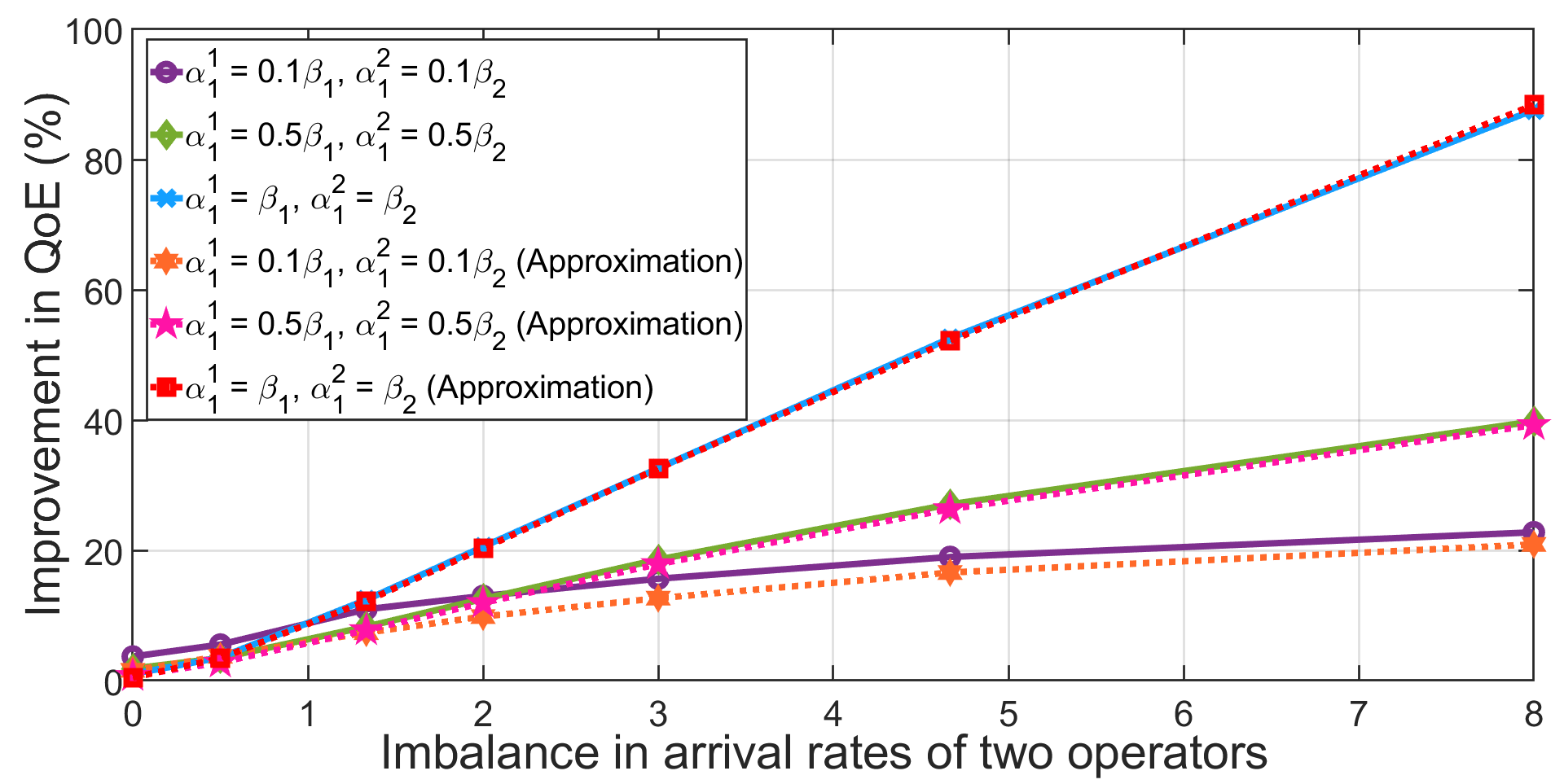}}
\caption{Performance comparison using a lower-complexity convex approximation}
\label{approx_b}
\end{figure}

Fig. \ref{approx_b} studies the performance of a lower-complexity variation of  our policy replacing $b_{r,n}^{i}(k)$ (involving a non-convex indicator function) with convex piece-wise linear approximation
$\min \left(1, Q_{r,n}^{i}(k)/Q_{min}\right).$
With this approximation, the optimization problem \eqref{optmzn_9a}-\eqref{optmzn_9b} turns into a convex optimization problem.
We see in Fig. \ref{approx_b} that the lower complexity variation's performance (shown using dashed lines) is comparable to that without the approximation (shown using solid lines).

\begin{figure}[htbp]
\centerline{\includegraphics[scale=0.17]{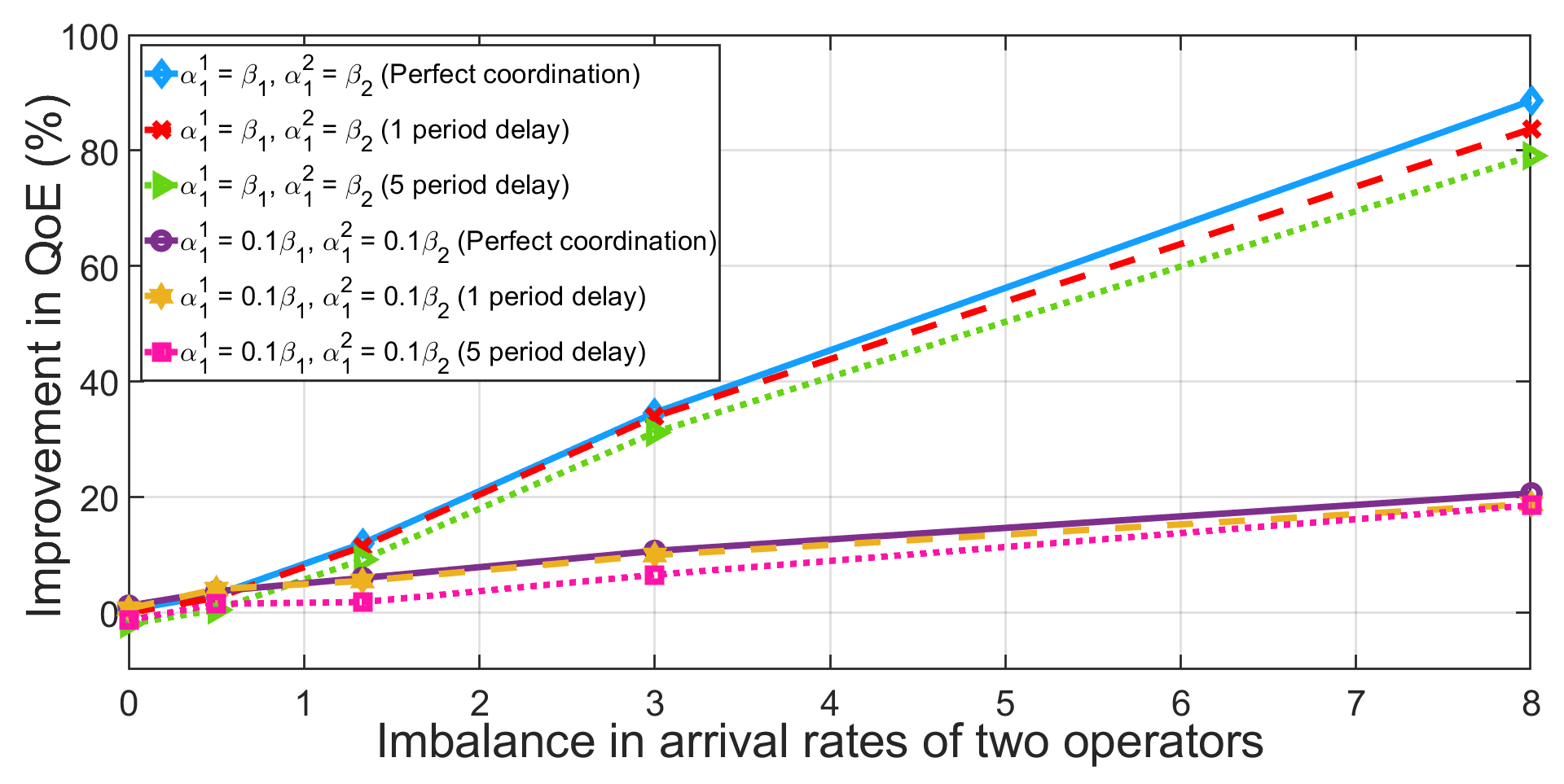}}
\caption{Effect of imperfect inter-operator coordination and delays in resource allocation}
\label{delay}
\end{figure}

In Fig. \ref{delay} we consider correlated packet arrivals and study the effect of imperfect inter-operator coordination/delay in resource allocation. Here, we use an AR1 process to generate the correlated packet arrivals and the correlation coefficient was set to 0.5. The results demonstrate that our policy maintains robust performance even under delays or latencies in the information exchange between operators.

\begin{figure}[htbp]
\centerline{\includegraphics[scale=0.17]{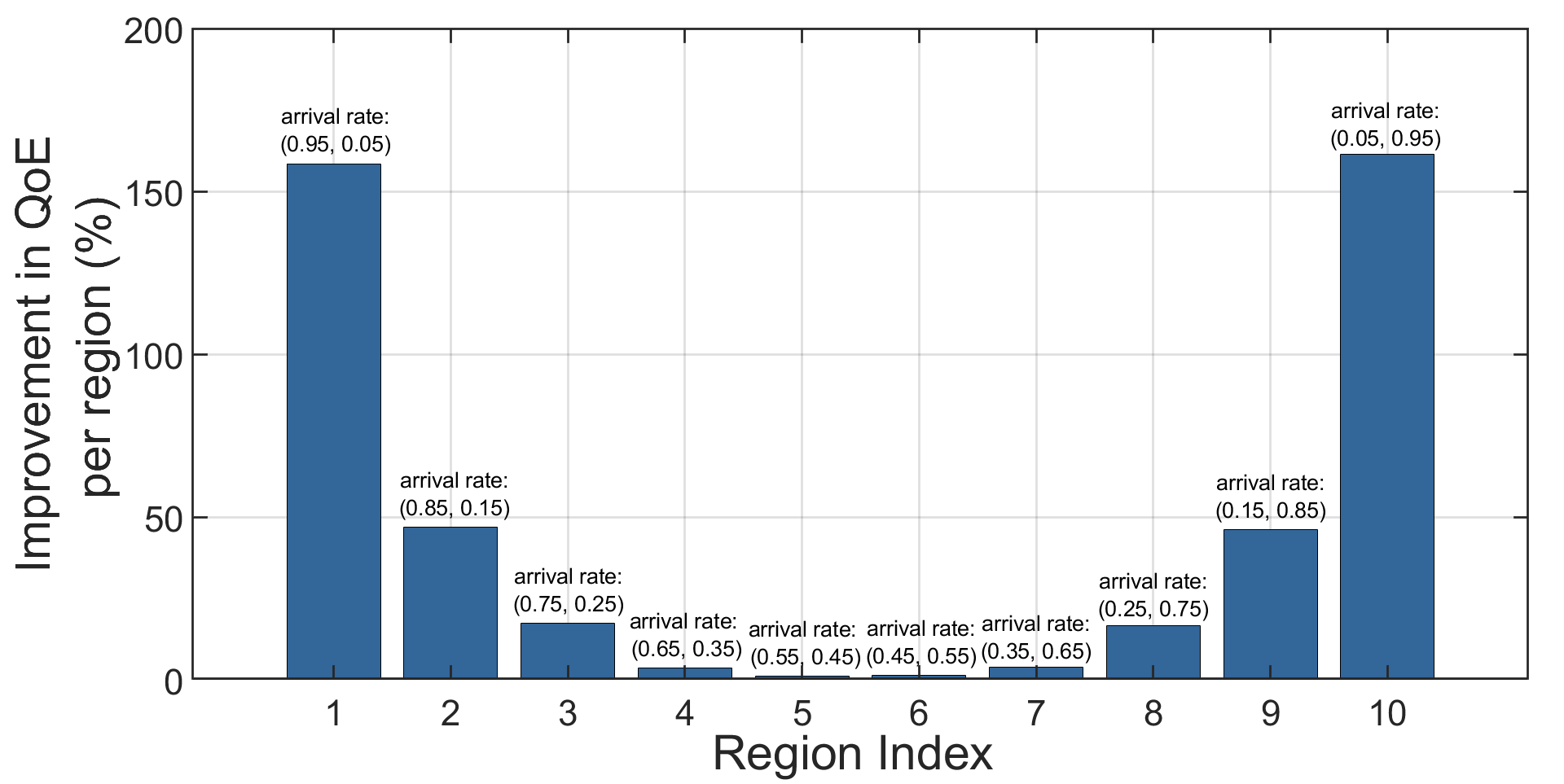}}
\caption{Scalability}
\label{scalability}
\end{figure}

Fig. \ref{scalability} illustrates the scalability of our policy when the number of regions increases. Here alone the average packet arrival rate of a client per period of one operator in a region is set to take values from \{0.05, 0.15,..., 0.95\}, while the average packet arrival rate of clients of the other operator in the same region is defined as 
one minus the corresponding value from the aforementioned set. Substantial QoE improvement per region is observed when there is a significant imbalance in arrival rates between the operators. Additionally, we note that with a large number of regions, clusters can be formed, allowing our policy to be effectively applied to each cluster.
 
\section{Conclusion} \label{sectionconclusion}
Our policy performing QoE-aware bandwidth sharing provides substantial QoE improvements (up to 90\%) for real-time traffic without requiring additional investment in spectrum, and these are also backed by strong  theoretical performance guarantees.
The amount of QoE improvement depends on parameters like traffic characteristics, channel conditions etc. 
In particular, the improvements are more when the arrival rates of different operators are imbalanced or when there is an imbalance in channel conditions of the operators. 

A useful extension of this work is a less dynamic solution for sharing, which does not require very frequent coordination between operators. It will also be interesting to consider other (e.g., monetary) incentive mechanisms for operators to incentivize sharing. The framework and performance results in this work act as useful reference points for such extensions.

\ifCLASSOPTIONcaptionsoff
  \newpage
\fi

\bibliographystyle{IEEEtran}
\bibliography{references.bib}

\begin{thebibliography}{10}
\providecommand{\url}[1]{#1}
\csname url@samestyle\endcsname
\providecommand{\newblock}{\relax}
\providecommand{\bibinfo}[2]{#2}
\providecommand{\BIBentrySTDinterwordspacing}{\spaceskip=0pt\relax}
\providecommand{\BIBentryALTinterwordstretchfactor}{4}
\providecommand{\BIBentryALTinterwordspacing}{\spaceskip=\fontdimen2\font plus
\BIBentryALTinterwordstretchfactor\fontdimen3\font minus \fontdimen4\font\relax}
\providecommand{\BIBforeignlanguage}[2]{{%
\expandafter\ifx\csname l@#1\endcsname\relax
\typeout{** WARNING: IEEEtran.bst: No hyphenation pattern has been}%
\typeout{** loaded for the language `#1'. Using the pattern for}%
\typeout{** the default language instead.}%
\else
\language=\csname l@#1\endcsname
\fi
#2}}
\providecommand{\BIBdecl}{\relax}
\BIBdecl

\bibitem{RM}
\BIBentryALTinterwordspacing
{Research and Markets}, ``Global video conferencing market analysis, forecast to 2023.'' [Online]. Available: \url{https://www.researchandmarkets.com/reports/4750113/global-video-conferencing-market-analysis}
\BIBentrySTDinterwordspacing

\bibitem{gmi}
\BIBentryALTinterwordspacing
{GMI Global Market Insights}, ``Video conferencing market size, growth trends, 2027 report.'' [Online]. Available: \url{https://www.gminsights.com/industry-analysis/video-conferencing-market}
\BIBentrySTDinterwordspacing

\bibitem{meti_research}
\BIBentryALTinterwordspacing
{Meticulous Research}, ``Live streaming market by component (platform, services), offering model \text{(B2B, B2C)}, streaming type (audio, video, game), vertical (media \& entertainment, education, sports \& gaming, government, fitness), and region - global forecast to 2028.'' [Online]. Available: \url{https://www.meticulousresearch.com/product/live-streaming-market-5225}
\BIBentrySTDinterwordspacing

\bibitem{3gpp_carrier}
\BIBentryALTinterwordspacing
3GPP, ``Carrier aggregation on mobile networks.'' [Online]. Available: \url{https://www.3gpp.org/technologies/carrier-aggregation-on-mobile-networks}
\BIBentrySTDinterwordspacing

\bibitem{ofcom}
\BIBentryALTinterwordspacing
Ofcom, ``The future role of spectrum sharing for mobile and wireless data services - \text{Licensed sharing, Wi-Fi}, and dynamic spectrum access.'' [Online]. Available: \url{https://www.ofcom.org.uk/consultations-and-statements/category-1/spectrum-sharing}
\BIBentrySTDinterwordspacing

\bibitem{nist}
\BIBentryALTinterwordspacing
{National Institute of Standards and Technology, US Department of Commerce}. Spectrum sharing. [Online]. Available: \url{https://www.nist.gov/advanced-communications/spectrum-sharing}
\BIBentrySTDinterwordspacing

\bibitem{cbrs}
\BIBentryALTinterwordspacing
\text{CBRS WlinForum Standards}, ``What role does the forum play?'' [Online]. Available: \url{https://cbrs.wirelessinnovation.org/about}
\BIBentrySTDinterwordspacing

\bibitem{deptindia}
\BIBentryALTinterwordspacing
{Department of Telecommunications, Government of India}, ``Spectrum sharing guidelines 2021.'' [Online]. Available: \url{https://dot.gov.in/spectrummanagement/spectrum-sharing-guidelines-2021}
\BIBentrySTDinterwordspacing

\bibitem{joseph2012jointly}
V.~Joseph and G.~de~Veciana, ``Jointly optimizing multi-user rate adaptation for video transport over wireless systems: Mean-fairness-variability tradeoffs,'' in \emph{2012 Proceedings IEEE INFOCOM}.\hskip 1em plus 0.5em minus 0.4em\relax IEEE, 2012, pp. 567--575.

\bibitem{cermak2011relationship}
G.~Cermak, M.~Pinson, and S.~Wolf, ``The relationship among video quality, screen resolution, and bit rate,'' \emph{IEEE transactions on broadcasting}, vol.~57, no.~2, pp. 258--262, 2011.

\bibitem{van2008traffic}
G.~Van~der Auwera, P.~T. David, and M.~Reisslein, ``Traffic characteristics of \text{H. 264/AVC} variable bit rate video,'' \emph{IEEE Communications Magazine}, vol.~46, no.~11, pp. 164--174, 2008.

\bibitem{wang2004video}
Z.~Wang, L.~Lu, and A.~C. Bovik, ``Video quality assessment based on structural distortion measurement,'' \emph{Signal processing: Image communication}, vol.~19, no.~2, pp. 121--132, 2004.

\bibitem{paris1999zero}
J.-F. P{\^a}ris, D.~D. Long, and P.~E. Mantey, ``Zero-delay broadcasting protocols for video-on-demand,'' in \emph{Proceedings of the seventh ACM international conference on Multimedia (Part 1)}, 1999, pp. 189--197.

\bibitem{palm1953methods}
C.~Palm, ``Methods of judging the annoyance caused by congestion,'' \emph{Tele}, vol.~4, no. 189208, pp. 4--5, 1953.

\bibitem{bar}
D.~Y. Barrer, ``Queuing with impatient customers and ordered service,'' \emph{Operations Research}, vol.~5, pp. 650--656, 1957.

\bibitem{kok}
A.~G.~D. Kok and H.~G. Tijms, ``A queueing system with impatient customers,'' \emph{Journal of Applied Probability}, vol.~22, pp. 688--696, 1985.

\bibitem{cosmetatos1985approximate}
G.~P. Cosmetatos and G.~P. Prastacos, ``An approximate analysis of the \text{D/M/1} queue with deterministic customer impatience,'' \emph{RAIRO-Operations Research}, vol.~19, no.~2, pp. 133--142, 1985.

\bibitem{boots1999multiserver}
N.~K. Boots and H.~Tijms, ``A multiserver queueing system with impatient customers,'' \emph{Management Science}, vol.~45, pp. 444--448, 1999.

\bibitem{leho1}
J.~Lehoczky, ``Real-time queueing theory,'' in \emph{17th IEEE Real-Time Systems Symposium}, 1996, pp. 186--195.

\bibitem{kruk}
{\L}.~Kruk, J.~Lehoczky, K.~Ramanan, and S.~Shreve, ``Heavy traffic analysis for edf queues with reneging,'' \emph{The Annals of Applied Probability}, vol.~21, no.~2, pp. 484--545, 2011.

\bibitem{lil}
B.~Aazhang, J.~Lilleberg, and G.~Middleton, ``Spectrum sharing in a cellular system,'' in \emph{Eighth IEEE International Symposium on Spread Spectrum Techniques and Applications - Programme and Book of Abstracts (IEEE Cat. No.04TH8738)}, 2004, pp. 355--359.

\bibitem{joshi}
S.~K. Joshi, K.~S. Manosha, M.~Codreanu, and M.~Latva-aho, ``Dynamic inter-operator spectrum sharing via \text{Lyapunov} optimization,'' \emph{IEEE Trans. on Wireless Communications}, vol.~16, no.~10, pp. 6365--6381, 2017.

\bibitem{ikami2020dynamic}
A.~Ikami, T.~Hayashi, and Y.~Amano, ``Dynamic channel allocation algorithm for spectrum sharing between different radio systems,'' in \emph{2020 IEEE 31st Annual International Symposium on Personal, Indoor and Mobile Radio Communications}.\hskip 1em plus 0.5em minus 0.4em\relax IEEE, 2020, pp. 1--6.

\bibitem{bennis2007inter}
M.~Bennis and J.~Lilleberg, ``Inter base station resource sharing and improving the overall efficiency of \text{B3G} systems,'' in \emph{2007 IEEE 66th Vehicular Technology Conference}.\hskip 1em plus 0.5em minus 0.4em\relax IEEE, 2007, pp. 1494--1498.

\bibitem{hou2009theory}
I.-H. Hou, V.~Borkar, and P.~R. Kumar, ``A theory of \text{QoS} for wireless,'' in \emph{IEEE INFOCOM 2009}, 2009, pp. 486--494.

\bibitem{hou2013scheduling}
I.-H. Hou, ``Scheduling heterogeneous real-time traffic over fading wireless channels,'' \emph{IEEE/ACM Transactions on Networking}, vol.~22, no.~5, pp. 1631--1644, 2013.

\bibitem{prk}
I.-H. Hou and P.~R. Kumar, ``Scheduling heterogeneous real-time traffic over fading wireless channels,'' in \emph{2010 Proceedings IEEE INFOCOM}, 2010, pp. 1--9.

\bibitem{lu2018age}
N.~Lu, B.~Ji, and B.~Li, ``Age-based scheduling: Improving data freshness for wireless real-time traffic,'' in \emph{Proceedings of the eighteenth ACM international symposium on mobile ad hoc networking and computing}, 2018, pp. 191--200.

\bibitem{tsanikidis2021power}
C.~Tsanikidis and J.~Ghaderi, ``On the power of randomization for scheduling real-time traffic in wireless networks,'' \emph{IEEE/ACM Transactions on Networking}, vol.~29, no.~4, pp. 1703--1716, 2021.

\bibitem{age_optimal_sun}
J.~Sun, L.~Wang, Z.~Jiang, S.~Zhou, and Z.~Niu, ``Age-optimal scheduling for heterogeneous traffic with timely throughput constraints,'' \emph{IEEE Journal on Selected Areas in Communications}, vol.~39, no.~5, pp. 1485--1498, 2021.

\bibitem{timely_qi}
Q.~Wang, C.~He, K.~Jaffrès-Runser, J.~Huang, and Y.~Xu, ``Timely-throughput optimal scheduling for wireless flows with deep reinforcement learning,'' in \emph{2022 IEEE/ACM 30th International Symposium on Quality of Service \text{(IWQoS)}}, 2022, pp. 1--11.

\bibitem{ICC_paper}
S.~A. George and V.~Joseph, ``Optimizing bandwidth sharing for real-time traffic in wireless networks,'' in \emph{ICC 2023 - IEEE International Conference on Communications}, 2023, pp. 3199--3204.

\bibitem{huang_joint_source}
J.~Huang, Z.~Li, M.~Chiang, and A.~K. Katsaggelos, ``Joint source adaptation and resource allocation for multi-user wireless video streaming,'' \emph{IEEE Transactions on Circuits and Systems for Video Technology}, vol.~18, no.~5, pp. 582--595, 2008.

\bibitem{fu2010systematic}
F.~Fu and M.~Van Der~Schaar, ``A systematic framework for dynamically optimizing multi-user wireless video transmission,'' \emph{IEEE Journal on Selected Areas in Communications}, vol.~28, no.~3, pp. 308--320, 2010.

\bibitem{webex}
\BIBentryALTinterwordspacing
{Webex-Help Center}, ``Troubleshoot \text{Webex} calling media quality in control hub.'' [Online]. Available: \url{https://help.webex.com/en-us/article/frj1efb/Troubleshoot-Webex-Calling-Media-Quality-in-Control-Hub}
\BIBentrySTDinterwordspacing

\bibitem{neely}
L.~Georgiadis, M.~Neely, and L.~Tassiulas, ``Resource allocation and cross-layer control in wireless networks,'' \emph{Foundations and Trends in Networking}, vol.~1, 01 2006.

\bibitem{minitool}
\BIBentryALTinterwordspacing
{MiniTool Video Converter}, ``What’s the best frame rate for 4k video? 60 fps or 50 fps?'' [Online]. Available: \url{https://videoconvert.minitool.com/video-converter/best-frame-rate-for-4k-video.html}
\BIBentrySTDinterwordspacing

\bibitem{DVDFab}
\BIBentryALTinterwordspacing
{DVDFab}, ``Best frame rate for 4k video: A beginner guide.'' [Online]. Available: \url{https://www.dvdfab.cn/resource/video/best-frame-rate-for-4k-video}
\BIBentrySTDinterwordspacing

\bibitem{hou2017wireless}
X.~Hou, Y.~Lu, and S.~Dey, ``Wireless \text{VR/AR} with edge/cloud computing,'' in \emph{2017 26th International Conference on Computer Communication and Networks (ICCCN)}.\hskip 1em plus 0.5em minus 0.4em\relax IEEE, 2017, pp. 1--8.

\bibitem{hazarika2023towards}
A.~Hazarika and M.~Rahmati, ``Towards an evolved immersive experience: Exploring \text{5G}-and beyond-enabled ultra-low-latency communications for augmented and virtual reality,'' \emph{Sensors}, vol.~23, no.~7, p. 3682, 2023.

\bibitem{itu911}
\text{Telecommunication Standardization Sector of ITU (ITU-T)}, ``Subjective audiovisual quality assessment methods for multimedia applications \text{(ITU-T Rec. P.911)},'' 1998.

\bibitem{centurylink}
\BIBentryALTinterwordspacing
{CenturyLink}, ``How to improve your gaming latency.'' [Online]. Available: \url{https://www.centurylink.com/home/help/internet/how-to-improve-gaming-latency.html}
\BIBentrySTDinterwordspacing

\bibitem{screenbeam}
\BIBentryALTinterwordspacing
{ScreenBeam}, ``How to reduce latency or lag in gaming.'' [Online]. Available: \url{https://www.screenbeam.com/wifihelp/wifibooster/how-to-reduce-latency-or-lag-in-gaming-2/}
\BIBentrySTDinterwordspacing

\bibitem{wiredshopper}
\BIBentryALTinterwordspacing
{WiredShopper}, ``What is a good latency for gaming?'' [Online]. Available: \url{https://thewiredshopper.com/what-is-a-good-latency-for-gaming/}
\BIBentrySTDinterwordspacing

\bibitem{haivision}
\BIBentryALTinterwordspacing
{Haivision}, ``Video latency.'' [Online]. Available: \url{https://www.haivision.com/glossary/video-latency/}
\BIBentrySTDinterwordspacing

\bibitem{adobe}
\BIBentryALTinterwordspacing
{Adobe}, ``Choosing the right video format.'' [Online]. Available: \url{https://www.adobe.com/in/creativecloud/video/discover/best-video-format.html}
\BIBentrySTDinterwordspacing

\bibitem{wiki}
\BIBentryALTinterwordspacing
{Wikipedia}, ``Video codec.'' [Online]. Available: \url{https://en.wikipedia.org/wiki/Video_codec}
\BIBentrySTDinterwordspacing

\bibitem{wowza}
\BIBentryALTinterwordspacing
{Wowza Media Systems}, ``Video codecs and encoding: Everything you should know.'' [Online]. Available: \url{https://www.wowza.com/blog/video-codecs-encoding}
\BIBentrySTDinterwordspacing

\bibitem{microsoft}
\BIBentryALTinterwordspacing
{Microsoft}, ``Real-time media calls and meetings with microsoft teams.'' [Online]. Available: \url{https://learn.microsoft.com/en-us/microsoftteams/platform/bots/calls-and-meetings/real-time-media-concepts}
\BIBentrySTDinterwordspacing

\bibitem{zoom}
\BIBentryALTinterwordspacing
{ Zoom Video Communications}, ``Conference room connector supported devices.'' [Online]. Available: \url{https://support.zoom.us/hc/en-us/articles/202445433-Conference-Room-Connector-supported-devices}
\BIBentrySTDinterwordspacing

\bibitem{webex_VC}
\BIBentryALTinterwordspacing
{Webex}, ``Known issues and limitations for webex video.'' [Online]. Available: \url{https://help.webex.com/en-us/article/gw8u4j/Known-issues-and-limitations-for-Webex-video}
\BIBentrySTDinterwordspacing

\bibitem{5GNR}
M.~Fuentes \emph{et~al.}, ``5g new radio evaluation against \text{IMT}-2020 key performance indicators,'' \emph{IEEE Access}, vol.~8, pp. 110\,880--110\,896, 2020.

\bibitem{coleago}
\BIBentryALTinterwordspacing
{ Coleago Consulting Ltd}, ``Estimating the mid-band spectrum needs in the 2025-2030 time frame.'' [Online]. Available: \url{https://www.gsma.com/spectrum/wp-content/uploads/2021/07/Estimating-Mid-Band-Spectrum-Needs.pdf}
\BIBentrySTDinterwordspacing

\bibitem{boris}
\BIBentryALTinterwordspacing
{Boris \text{FX}}. \text{SD} vs \text{HD} resolution: What are the main differences? [Online]. Available: \url{https://borisfx.com/blog/sd-vs-hd-resolution-the-main-differences/}
\BIBentrySTDinterwordspacing

\bibitem{resi}
\BIBentryALTinterwordspacing
{Resi Media \text{LLC}}. What are video resolutions? [Online]. Available: \url{https://resi.io/glossary/video-resolutions/}
\BIBentrySTDinterwordspacing

\bibitem{minBRgoogle}
\BIBentryALTinterwordspacing
{Google}. Prepare your network for meet meetings \& live streams. [Online]. Available: \url{https://support.google.com/a/answer/1279090?hl=en#zippy=%2Cstep-review-bandwidth-requirements%2Cstep-set-up-outbound-ports-for-media-traffic%2Cstep-allow-access-to-uniform-resource-identifiers-uris%2Cstep-allow-access-to-google-ip-address-ranges-for-audio-and-video}
\BIBentrySTDinterwordspacing

\bibitem{dacast}
\BIBentryALTinterwordspacing
{Dacast}. Bitrate vs. resolution for professional broadcasting. [Online]. Available: \url{https://www.dacast.com/blog/bitrate-vs-resolution/#:~:text=Bitrate%20and%20resolution%20are%20two%20important%20measures%20which%20go%20together,difference%20to%20your%20video%20quality.}
\BIBentrySTDinterwordspacing

\bibitem{papoulis2002probability}
A.~Papoulis and S.~U. Pillai, \emph{Probability, random variables, and stochastic processes}.\hskip 1em plus 0.5em minus 0.4em\relax Tata McGraw-Hill Education, 2002.

\bibitem{neelythesis}
M.~J. Neely, ``Dynamic power allocation and routing for satellite and wireless networks with time varying channels,'' Ph.D. dissertation, Massachusetts Institute of Technology, 2003.

\end{thebibliography}

\result

\end{document}